\documentclass[a4paper,11pt]{article}
\pdfoutput=1 

\usepackage{jheppub} 

\usepackage[T1]{fontenc} 

\usepackage{lipsum}
\usepackage{graphicx}
\usepackage{verbatim}
\usepackage{subfigure}
\usepackage{amsmath,amssymb,amsfonts,eufrak,mathrsfs}
\usepackage{bm}
\usepackage{color}

\usepackage{lineno}

\usepackage{dcolumn}
\usepackage{bm}
\usepackage{amsmath,eufrak,mathrsfs}

\newcommand \kd  {\delta}
\newcommand \ra  {\rightarrow}

\newcommand \w  {\omega}

\newcommand \A {\alpha}

\newcommand \lc {\langle}
\newcommand \rc {\rangle}
\newcommand \prt {\partial}

\newcommand \sg {\sigma}

\newcommand \bvec{\left( \begin{array}{c} }
\newcommand \evec{\end{array} \right)}

\newcommand \bea{\begin{eqnarray} }
\newcommand \eea{\end{eqnarray} } 
\newcommand \nn {\nonumber}
\newcommand {\be} {\begin{equation}}
\newcommand {\ee} {\end{equation}}
\newcommand {\epem} {$e^+ e^-$}

\setrunninglinenumbers
\title{\boldmath The medium modification of high-virtuality partons}

\author[a,b]{Shanshan Cao}
\author[a]{Chathuranga Sirimanna}
\author[a]{Abhijit Majumder}

\affiliation[a]{Department of Physics and Astronomy, Wayne State University, Detroit, MI 48201}
\affiliation[b]{Institute of Frontier and Interdisciplinary Science, Shandong University, Qingdao, Shandong 266237, China}

\emailAdd{shanshan.cao@sdu.edu.cn}
\emailAdd{chathuranga.sirimanna@wayne.edu}
\emailAdd{majumder@wayne.edu}

\abstract{The modification of the hard core of jets in a dense QCD medium is studied. In particular, we consider partons which possess a virtuality somewhat larger than the multiple scattering scale of the medium ($\hat{q} \tau$, where $\hat{q}$ is the transverse broadening jet transport coefficient, and $\tau$ is the formation length of a particular emission). We delineate the region of parameter space where the higher-twist approach is applicable, and derive the in-medium DGLAP evolution equation. We study a region in parameter space where this is the dominant mechanism of energy loss. We argue that such a regime is pervasive in most cases of jets in $A$-$A$ and future $e$-$A$ collisions, and controls the modification of the hard core of jets and the leading single particle spectrum at high transverse momentum ($p_\mathrm{T}$).}

\begin{document} 
\maketitle
\flushbottom


\section{Introduction}
\label{sec:introduction}
Almost 20 years past the discovery of jet quenching in the suppression of high transverse momentum (high-$p_\mathrm{T}$) hadrons at the Relativistic Heavy-Ion Collider (RHIC) at Brookhaven National Laboratory (BNL)~\cite{Adcox:2001jp,Adler:2002tq}, 
and a decade past the observation of modification of reconstructed jets at the Large Hadron Collider (LHC) at CERN~\cite{Chatrchyan:2011sx,Aad:2010bu}, there is a measure of consensus regarding the overarching mechanism among the practitioners of the field. It is now widely held that jets are multi-scale objects~\cite{Cao:2020wlm,Cao:2017zih,Majumder:2014vpa,Majumder:2010qh}, that start as a single parton with an off-shellness or virtuality $\mu$ far above the ambient medium scale of temperature $T$, or Debye mass $m_\mathrm{D} \simeq gT \sim T$, in a heavy-ion collision, and continue to lose this off-shellness in a cascade of partonic splits, that take place over an extended lifetime.

Within this cascade, it is also generally accepted that there are at least two distinct stages~\cite{Caucal:2018dla,Armesto:2011ir,CasalderreySolana:2012ef,Majumder:2014vpa}: A higher virtuality stage (HVS), where the off-shellness of a given parton is greater than that generated by multiple scattering (MS) in the medium, i.e. $\mu^2 \gnsim \mu^2_{MS} \sim \hat{q}\tau$, where $\tau$ is the lifetime of the parton, and 
$\hat{q}$ is the transverse momentum transport coefficient defined for a massless parton as,
\bea
\hat{q} = \frac{ \lc k_\perp ^2 \rc_L }{L} \equiv \frac{ \lc k_\perp ^2 \rc_{L^0} }{ L^0 } \equiv \frac{\lc k_\perp^2 \rc_{L^-} }{L^-}, \label{qhat-def}
\eea
the momentum squared per unit length (or time $L^0 = L$, or light-cone length $L^- = [ L^0 - L_z ]/\sqrt{2} $) gained by a single parton traversing a length $L$, transverse to its original direction (designated as the negative $z$-axis). This higher virtuality stage is often referred to as the radiation stage~\cite{Majumder:2014vpa} or Dokshitzer-Gribov-Lipatov-Altarelli-Parisi (DGLAP) stage~\cite{Gribov:1972ri,Gribov:1972rt,Altarelli:1977zs,Dokshitzer:1977sg}. In the higher virtuality stage, the frequency of hard emissions outweighs that of hard scatterings at a similar scale, and the effect of these emissions can be calculated either by using the DGLAP equation or on the basis of event generators based on this equation.
This is followed by a lower virtuality stage (LVS), where the virtuality of the parton remains approximately at the scale generated by multiple scatterings 
$\mu^2_{MS} \sim \hat{q} \tau$. This is often referred to as the scattering stage or the Baier-Dokshitzer-Mueller-Peigne-Schiff (BDMPS) stage~\cite{Baier:1996sk,Baier:1996kr} or the Arnold-Moore-Yaffe (AMY) stage~\cite{Arnold:2001ms,Arnold:2001ba,Arnold:2002ja}. 
In the lower virtuality stage, the hard parton endures multiple scattering, with each scattering leading to a shift in the scale of $\mu^2_{SS} \sim \hat{q} \lambda$ 
($\lambda$ is the mean free path of a jet parton, the subscript $SS$ refers to single scattering) prior to an emission at the scale $\mu^2_{MS} = \hat{q} \tau$. 
If a parton undergoes $\lc n \rc$ scatterings on average, then $\tau = \lc n \rc \lambda$, yielding $\mu^2_{MS} = \lc n \rc \mu^2_{SS}$. 

In both these stages, the higher virtuality stage and the lower virtuality stage, the leading partons under consideration carry a substantial fraction of the energy of the jet. The energy of these partons are thus considerably high: $E \gg \mu$, and the scale $\mu$ is perturbatively large i.e., $\mu \gg T \gtrsim \Lambda_\mathrm{QCD}$.
(In this Paper, we will limit ourselves to the factual consideration that the temperatures reached in heavy-ion collisions at RHIC and LHC are of the order of the QCD non-perturbative scale and the preponderance of the quenching takes place at temperatures above this scale i.e., $T \gtrsim \Lambda_\mathrm{QCD}$.) In the lower virtuality stage, the scale fluctuates from emission to emission, but on average remains at the scale $\mu^2 _{MS}= \hat{q}\tau$, whereas in the higher virtuality stage, the scale drops with each emission, down to the medium generated scale. Thus, while almost all emissions in the lower virtuality stage are medium induced, in the higher virtuality stage, some of the emissions are vacuum like and some can be affected by scattering in the medium.

In several recent papers, it has been proposed, that the higher virtuality stage is entirely vacuum like~\cite{Caucal:2018dla}, meaning that there is no medium modification to the multiple emissions of this stage. A related argument states that the scale (virtuality/off-shellness) or $\mu^2$ of the virtual parton in the higher virtuality stage is so much larger than the medium scale $\mu^2_{MS} \sim \hat{q} \tau$~\cite{Armesto:2011ir,CasalderreySolana:2012ef}, that it cannot be resolved by partons in the medium, leading to no medium induced emission. This is often referred to as the \emph{coherence} effect in jet quenching. In this Paper, we will demonstrate that while a majority of the emissions in the higher virtuality stage are indeed vacuum like, the probability of scatterings and the resulting mean number of scatterings in the higher virtuality stage is non-negligible compared to the number of scatterings in the lower virtuality stage. Since each scattering in the higher virtuality stage takes place at a much larger scale than in the lower virtuality stage, its eventual effect on the evolution of the jet, and in particular the hard core of the jet, is in fact considerable.

The remainder of the paper is organized as follows, in Sect.~\ref{sec:par} we will present simple parametric estimates for virtuality evolution and the evolution of the ambient medium scale 
$\mu_{MS}$ for both static and dynamically expanding media. We will demonstrate that, while in static media the higher virtuality stage is indeed negligible compared to the lower virtuality stage for all jet observables (for media that are considerably long and for jet energies that are large enough for leading partons to decay beyond this length), in expanding media the higher virtuality stage lingers for a time comparable to the lifetime of the medium and thus has a considerable effect on the core of the jet. 
In Sect.~\ref{sec:vacscat}, we transition to more realistic estimates, using a real fluid-dynamical medium, and jets modeled using a two stage process. Ignoring any medium modification of the higher virtuality stage, we document the amount of time spent by the leading parton in the higher virtuality stage and the lower virtuality stage as a function of the parton's energy. It is demonstrated that as the energy goes above 150-200~GeV, which is very common at current LHC experiments, the time spent in the higher virtuality stage exceeds that in the lower virtuality stage. Due to this large time, the number of expected scatterings in the higher virtuality stage also exceed those in the lower virtuality stage, as a result, the picture of no interaction in the higher virtuality stage is rendered invalid. In Sect.~\ref{sec:NLO}, we revisit the calculation of the medium modified DGLAP evolution equation. A review of the published literature indicates that the medium modified DGLAP evolution equation has never been derived or expressed in full detail; this is carried out in Sect.~\ref{sec:NLO}. This allows us to precisely define what we mean by the effect of the medium being a perturbative correction to the process of vacuum emission at large virtuality. In Sect.~\ref{sec:med}, a Monte-Carlo event generator based on the derived DGLAP evolution equation is used to model the higher virtuality stage, in conjunction with a Boltzmann transport simulation for the lower virtuality stage. The simulation tends to show a slowing down of the rate at which the hard parton sheds virtuality and thus extends the amount of time spent by the leading parton in the higher virtuality stage. This further increases the modification of the leading parton in the higher virtuality stage. Our summary is presented in Sect.~\ref{sec:summary}.


\section{Parametric Estimates}
\label{sec:par}

Imagine a static cylindrical container of length $L$, and radius $R$ containing a Quark Gluon Plasma (QGP), i.e., a hot strongly interacting medium where quarks and gluons are deconfined, at a fixed temperature $T \gtrsim 0.2$~GeV. A single parton (quark or gluon) with energy $E$ and maximum virtuality $Q \lesssim E$ is created on the axis of the cylinder and travels radially outward. 
Imagine the cylinder is oriented along the beam axis of a heavy ion collision, and the radial direction corresponds to a direction transverse to the collision, i.e., $E \simeq p_\mathrm{T}$.

The parton will begin to split and decay into a cascade of partons with progressively lower virtuality. 
In this section, we will offer parametric estimates of the virtuality of the leading parton in the shower, as a function of time. Perturbative QCD splits tend to be rather ``undemocratic'', i.e., in a typical split, momentum fractions tend to be quite unequally distributed, and thus a large fraction of the energy of the jet tends to lie within one leading parton. In the remainder of this paper, we will be concerned with the status of this leading parton, and the leading hadron that emanates from a fragmentation of this leading parton.

As a starting point, let us assume that the medium does not interact with this developing shower, i.e., we are considering a vacuum like showering process taking place inside the medium. The successive splits can be reliably calculated using perturbative QCD (pQCD), until the virtuality of each parton reaches the scale $Q_0 \sim 1$~GeV. At this point, the leading parton will carry a fraction $z E$ of the energy of the original parton with $z \lesssim 1$. 
The lifetime of a parton with off-shellness or virtuality $\mu$, and energy $\w = z E$ is given as, 
\bea
\tau \simeq \frac{2\w}{\mu^2} = \frac{2 z E}{\mu^2} . \label{one_formation_length}
\eea
The above formula is an underestimate of the formation time. 
In reality, from the perspective of the leading parton, the total time spent in the medium can be decomposed into a series of formation times, each associated with the separation of an emitted parton from 
the emitting leading parton. Thus the time taken to emit $N$ partons in vacuum is, 
\bea
\tau_\mathrm{total} =  \sum_{i=1}^{N} \frac{2 z_1 z_2 \ldots z_{i-1} E   }{ \Delta \mu^2_i } . \label{time_for_N_emissions}
\eea 
In the above formula, the hard parton starts with energy $E$ and virtuality $\mu_0 = Q$, and loses both in successive emissions, with each emission exacting a drop in virtuality by $\Delta \mu_j^2 = \mu_{j-1}^2 - \mu_j^2$ ($\mu_0 = Q$) and a drop in momentum fraction by $z_j$.

In the limit that $\Delta \mu_j^2 \sim \mu_{j-1}^2$ for large drops in virtuality, one can invert Eq.~\eqref{one_formation_length} and obtain the virtuality of a parton at a time $\tau$ (having started with a virtuality of the order of the hard scale $\mu \sim Q$). 
This is easily approximated as, 
\bea
\mu^2(\tau) \gtrsim  \frac{2E}{\tau}. \label{mu_as_func_of_tau_single_emission}
\eea
This is typically an underestimate, the more accurate expression would be obtained by determining $N$ in Eq.~\eqref{time_for_N_emissions} such that $Q^2 - \sum_{i=1}^N \Delta \mu^2_i  = \mu^2$, 
and then  determining the total time by inverting Eq.~\eqref{time_for_N_emissions}. 
No doubt the actual number $N$ as well as the values of $\Delta \mu^2_j$ will vary from event to event, 
and thus the time to reach a given $\mu$ can either be determined event-by-event or in the average over the ensemble of events. 
In either case it will be longer than the lower estimate in Eq.~\eqref{one_formation_length}. Hence, the virtuality at a time $\tau$ will be larger than the lower estimate in Eq.~\eqref{mu_as_func_of_tau_single_emission}.

The higher virtuality stage starts with the production of the hard parton and continues till the parton reaches the scale where scattering from the medium can maintain the virtuality of the parton at the 
medium induced scale $Q_0^2 = \hat{q} \tau$. In a static medium, $\hat{q}$ is constant and thus $\tau_{Q_0}$ can be obtained as, 
\bea
Q_0^2 =  \frac{2 E}{\tau_{Q_0}} = \hat{q} \tau_{Q_0}, \,\,\,\, \Rightarrow \,\,\,\, \tau_{Q_0} = \sqrt{ \frac{ 2 E }{\hat{q}} }.  \label{tau_0_static_medium}
\eea
As would be expected, lower energy partons transition from the higher virtuality stage to the lower virtuality stage earlier than highier energy or leading partons. 
It should be pointed out again, that this is a lower estimate for the time at which the parton transitions from the higher virtuality stage to the lower virtuality stage, as this is usually attained via a series of sequential emissions.

If the radius of the cylindrical medium is such that $R \gg \tau_{Q_0} $ (assuming $c=1$), the predominant mechanism for jet modification, even for the leading parton, is the multiple scattering regime of the lower virtuality stage. 
In this limit of a large static medium, one could ignore any modification in the higher virtuality stage and only focus on the lower virtuality stage. Such a large static medium underlies the assumptions in the earlier papers highlighting the multiple 
scattering lower virtuality stage~\cite{Baier:1996sk,Baier:1996kr,Arnold:2001ms,Arnold:2001ba,Arnold:2002ja}.

A dynamically decaying medium tends to change this picture considerably. 
Let us now, once again, imagine that the medium is in the form of a cylinder of radius $R$, whose length $L$ increases at nearly the speed of light (Bjorken expansion~\cite{Bjorken:1982qr}).
We assume that $\hat{q}$ is uniform throughout the medium at a time $t_0$ (e.g., the thermalization time $t_0 \sim 1$~fm/c). We again consider hard partons that start on the axis with momentum 
in the transverse direction to the axis. Given the Bjorken expansion, the value of $\hat{q}$ at any time $\tau$ is given as, 
\bea
\hat{q} (\tau) = \hat{q}_0 \frac{t_0}{\tau}, \,\,\,\, {\rm with } \,\,\,\, \hat{q} = 0 \,\,\,\, {\rm for} \,\,\,\, \tau < t_0.
\eea

In such a medium, the hard leading parton would transition from the higher virtuality stage to the lower virtuality stage at $\tau_{Q_0}$ given as, 
\bea
Q_0^2 = \frac{2 E}{ \tau_{Q_0}} = \hat{q}_0 \frac{t_0}{\tau_{Q_0}} \tau_{Q_0}, \,\,\,\,  \Rightarrow \,\,\,\, \tau_{Q_0} = \frac{2E}{\hat{q}_0 t_0}. \label{tau_0_Bjorken_medium}
\eea
The transition time $\tau_{Q_0}$ in Eq.~\eqref{tau_0_Bjorken_medium} above, is much longer than the estimate for a static medium in Eq.~\eqref{tau_0_static_medium}.
In fact, partons with an energy $E > {\hat{q}_0 t_0 R }/{2}$, never transition from the higher virtuality stage to the lower virtuality stage. 
In a realistic heavy-ion collision, one dimensional Bjorken expansion, 
causing a $1/\tau$ behavior in density dependent quantities, 
is only expected to occur at early times. 
At late times, the expansion may indeed be three dimensional, leading to densities depleting as $1/\tau^3$. 
Also the density, and by extension $\hat{q} (\vec{x},\tau)$ is not uniform in space and time, it tends to be highest 
at the center of the colliding system and drops to the edges. Also while most jets start from the center and travel outward, several jets are formed away from the center and may travel through the densest part of the medium. 
To consider the effect of all these varying factors will require a realistic Monte-Carlo simulation, that simulates both the collision of the two nuclei, followed by the hydrodynamic expansion, and the development of a jet shower within this medium. This is presented in the subsequent sections.

In Sect.~\ref{sec:vacscat}, we present a more realistic version of the parametric estimate outlined above, 
where the jet shower develops in the medium, but does not interact with it, i.e., it develops as a vacuum shower. 
We find, as expected, that as the energy of the leading parton increases, it transitions from the higher virtuality stage to the lower virtuality stage at a later time in the evolution of the collision. 
In Sect.~\ref{sec:NLO}, we re-derive the theory of single scattering induced modification to the parton splitting kernel, 
as is appropriate for the higher virtuality stage. 
This kernel is then applied in Sect.~\ref{sec:med} to calculate the transition time between 
the higher virtuality stage and the lower virtuality stage, as experienced by the leading parton. 
The inclusion of the medium modified kernel tends to slow down the rate at which virtuality is lost by the leading parton. This is to be expected. 
Note that in the lower virtuality stage, scattering in the medium, effectively holds the virtuality at the scale $\mu_{MS}^2 = \hat{q} \tau$. 
Thus at scales higher than $\mu^2_{MS}$, scatterings tend to slow down the rate of drop of virtuality. This is described in detail in Sect.~\ref{sec:med}.


\section{Scatterings within vacuum DGLAP and transport stages}
\label{sec:vacscat}

A high energy parton produced in a hard interaction in a nucleus-nucleus collision tends to have high virtuality (produced far off mass shell). It evolves back to almost on-shell with successive splittings, typically described by the DGLAP evolution equation. With the presence of a thermal medium, this virtuality-ordered parton shower is expected to stop when the parton scale is comparable to that of the medium generated hard scale  $Q_0^2 \sim \hat{q}\tau$.

After reaching $Q_0^2$, the subsequent parton evolution is usually described by transport equations that include its elastic and inelastic scatterings through the medium. Whether the evolution of a given parton is dominated by the high virtuality (DGLAP) or low virtuality (transport) process relies on the medium properties, such as its length and density evolution profiles. In this section, we utilize a realistic hydrodynamic medium to evaluate the time a parton spends in the DGLAP versus the transport stages. With this time information, the numbers of scatterings within these two stages are also estimated. We will show that an energetic parton usually spends sufficiently long time within the QGP before its virtuality drops to $Q_0^2$, which allows it to scatter with the medium at least once during the DGLAP stage. 

To calculate the time a highly virtual parton spends to approach $Q_0^2$, we utilize the \textsc{Matter} event generator~\cite{Majumder:2013re,Cao:2017qpx} to simulate parton showers. This event generator is constructed based on the Sudakov formalism that simulates the DGLAP evolution with the Monte-Carlo method~\cite{Hoche:2014rga}. For an energetic parton produced at a point $r$ with a forward light-cone momentum $p^{-} = (p^0 + \hat{n}\cdot \vec{p})/\sqrt{2}$ ($\hat{n}=\vec{p}/|\vec{p}|$ denotes the direction of the parton, and we pick the negative light-cone component to be the large component by convention based on Deep Inelastic Scattering), the following Sudakov form factor gives the probability of no splitting between virtuality scales $t$ ($t = \mu^{2}$) and $t_\mathrm{max}$:
\begin{equation}
\label{eq:sudakov}
\Delta(t_\mathrm{max},t)=\exp\left[- \int\limits_t^{t_\mathrm{max}}\frac{d\tilde{t}}{\tilde{t}}\frac{\alpha_\mathrm{s}(\tilde{t})}{2\pi}\int\limits_{z_\mathrm{c}}^{1-z_\mathrm{c}} dy P(y,\tilde{t})\right].
\end{equation}
Here, $z_\mathrm{c}=t_\mathrm{min}/\tilde{t}$ is taken from the kinematic constraints of splitting with the minimum allowed virtuality, and is set as $t_\mathrm{min}=1~\mathrm{GeV}^2$. In this section, we use the vacuum splitting function $P$ in the high virtuality (DGLAP) stage as suggested by Ref.~\cite{Caucal:2018dla}. This calculation will be repeated in Sect.~\ref{sec:med} with $P$ including both vacuum and medium-induced parts, after this formalism is derived in Sect.~\ref{sec:NLO}.

With this setup, splitting of a given parton is allowed if $r>\Delta(t_\mathrm{max},t_\mathrm{min})$ is satisfied, where $r$ is  a random number uniformly sampled within $(0,1)$. The actual virtuality scale $t$ of the parton at which it splits is obtained by solving $r=\Delta(t_\mathrm{max},t)$, and the momentum fraction $y$ shared by the two daughter partons is determined by the splitting function $P(y)$. On the other hand, we treat the parton as stable and set $t=t_\mathrm{min}$ if $r\le\Delta(t_\mathrm{max},t_\mathrm{min})$. 
Each parton starts with the maximum possible virtuality $t_\mathrm{max}=E^2$, where $E$ is the energy of the parton immediately after the initial hard scattering. After the first splitting, $y^2 t$ and $(1-y)^2 t$ serve as the new upper limits $t_1^\mathrm{max}$ and $t_2^\mathrm{max}$ for the two daughter partons, from which their actual virtualities $t_1$ and $t_2$ are determined. The transverse momentum of the two daughter partons with respect to their parent is then calculated with
\begin{equation}
\label{eq:kT}
k_\perp^2=y(1-y)t-(1-y)t_1-yt_2.
\end{equation}
The location of the splitting, i.e., where the two daughter partons are produced, is calculated via $\vec{r}+\hat{n}\zeta$, where $\zeta$ is the actual splitting time of the parent parton (since its production) that is sampled via a Gaussian distribution with a mean value of $\tau^- = 2 p^- / t$ in light-cone coordinates:
\begin{equation}
\label{eq:tau}
\rho(\zeta^-)=\frac{2}{\tau^-\pi}\exp\left[-\left(\frac{\zeta^-}{\tau^-\sqrt{\pi}}\right)^2\right].
\end{equation}

This splitting process is iterated to generate a virtuality-ordered parton shower starting from a single parton with an initial $t_\mathrm{max} = E^2$, to a number of final-state partons at or below a switching scale $Q_0^2$, which is set as $t_\mathrm{min}=1~\mathrm{GeV}^2$ in vacuum, or set to $\max (\hat{q}\tau, t_\mathrm{min})$ in a thermal medium.  

This \textsc{Matter} event generator is coupled to a hydrodynamic model to simulate parton showers in realistic heavy-ion collisions. In this work, we use the (2+1)-dimensional viscous hydrodynamic model \textsc{Vishnew}~\cite{Song:2007fn,Song:2007ux,Qiu:2011hf} to generate a QGP fireball for central (0-5\%) Pb-Pb collisions at $\sqrt{s_\mathrm{NN}}=2.76$~TeV with smooth initial entropy distribution given by the Monte-Carlo (MC) Glauber model. The starting time of the QGP evolution ($\tau_0=0.6$~fm) and the specific shear viscosity ($\eta/s$=0.08) have been tuned to describe the spectra of soft hadrons emitted from the QGP at both RHIC and the LHC. The hydrodynamic model provides the spacetime evolution profiles of the local temperature ($T$), entropy density ($s$) and flow velocity ($u$) of the QGP, with which we estimate the local scattering rate $\Gamma$ (number of scatterings per unit length) of the jet parton and its transport coefficient $\hat{q}$. In this section, we assume the jet parton scatters with a massless thermal parton perturbatively at leading order. This yields~\cite{He:2015pra,Qin:2009gw}
\begin{align}
\label{eq:rate}
&\Gamma_\mathrm{local} \approx 42 C_R \zeta(3)\frac{\alpha_\mathrm{s}^2 T^3}{\pi \mu_\mathrm{D}^2},\\
\label{eq:qhat}
&\hat{q}_\mathrm{local} \approx 42 C_R \zeta(3) \frac{\alpha_\mathrm{s}^2 T^3}{\pi} \ln \left(\frac{5.7 ET}{4\mu_\mathrm{D}^2}\right),
\end{align}
where $C_R$ and $E$ represent the color factor and the energy of a given parton respectively, $\mu_\mathrm{D} = 6\pi\alpha_\mathrm{s}T^2$ is the Debye screen mass with $\alpha_\mathrm{s}$ as the strong coupling constant. Equations~(\ref{eq:rate}) and (\ref{eq:qhat}) are evaluated in the fluid rest frame. Effects of the local velocity of the expanding medium can be taken into account by utilizing the rescaled scattering rate $\Gamma=\Gamma_\mathrm{local}\cdot p^\mu u_\mu/p^0$ and jet transport coefficient $\hat{q}=\hat{q}_\mathrm{local}\cdot p^\mu u_\mu/p^0$~\cite{Baier:2006pt}.

\begin{figure}[tbp]
        \centering
                \includegraphics[width=0.55\linewidth]{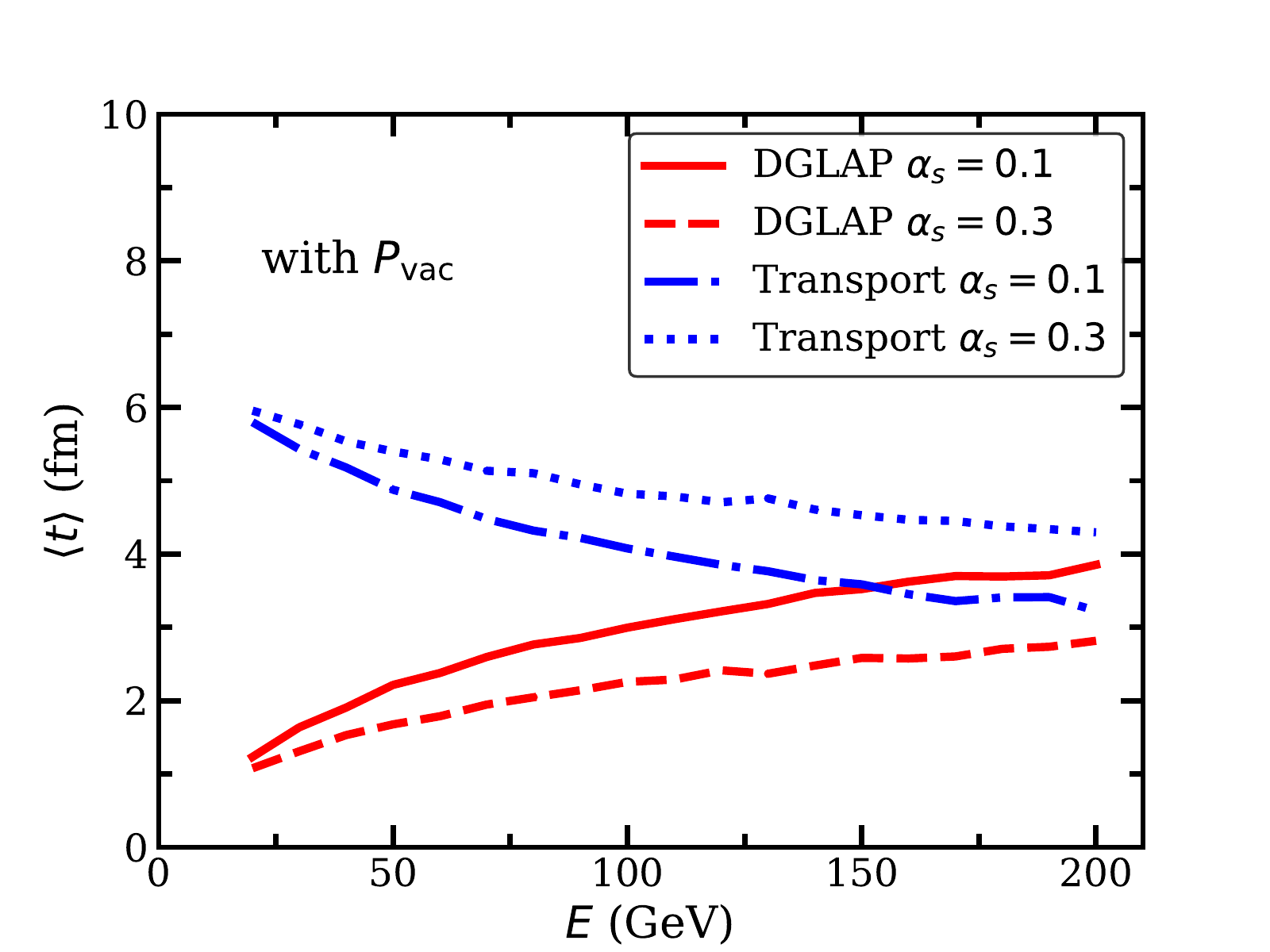}
        \caption{(Color online) The time spent by leading partons within high virtuality (DGLAP) versus low virtuality (transport) stages, with vacuum splitting function used in \textsc{Matter} simulation.}
        \label{fig:plot-average_t_g_vac}
\end{figure}

Within this framework, we track the leading parton within each jet as it propagates through the medium. Its total time spent in the DGLAP stage is obtained by summing over the formation lengths ($\zeta$'s) of the successive splittings in \textsc{Matter}, before its virtuality drops below $Q_0^2 = \hat{q}\tau$. After reaching $Q_0^2$, we let the parton continue streaming through the medium until the local temperature of the medium falls below the critical value $T_\mathrm{c}=165$~MeV. This second half of evolution is treated as the transport stage and will be simulated with the \textsc{Lbt} model in Sect.~\ref{sec:med} when we calculate realistic observables.

In Fig.~\ref{fig:plot-average_t_g_vac}, we compare the average times (and by extension, lengths) a leading parton spends within the high virtuality (DGLAP) versus the low virtuality (transport) stages, as functions of the initial parton energy. The average is taken over initial partons that are sampled at different locations inside the medium based on the MC Glauber model. Two different $\alpha_\mathrm{s}$ values are applied and compared. One observes that as its energy increases, the parton spends a longer time in the high virtuality (DGLAP) stage, prior to approaching $Q_0$, because its initial virtuality increases with energy. Since in this section we assume partons in the high virtuality stage do not interact with the medium and use the vacuum splitting function in \textsc{Matter}, changing $\alpha_\mathrm{s}$ does not affect the rate at which partons lose virtuality. However, increasing $\alpha_\mathrm{s}$ raises $\hat{q}$ and $Q_0$, thus terminates the DGLAP showers earlier. 
Lower energy partons spend a shorter time in the high virtuality (DGLAP) stage and therefore spend a longer time in the low virtuality (transport) stage. On the other hand, a 200~GeV gluon spends comparable times in these two stages. The total time a parton spends inside the medium is around 7~fm regardless of its energy, consistent with the average lifetime of QGP medium produced in 2.76~ATeV central Pb-Pb collisions. Note that energetic partons spend a considerable amount of time inside the medium within the high virtuality (DGLAP) stage, indicating the necessity to incorporate and study medium modification with DGLAP evolution for these highly virtual partons.

\begin{figure}[tbp]
        \centering
                \includegraphics[width=0.55\linewidth]{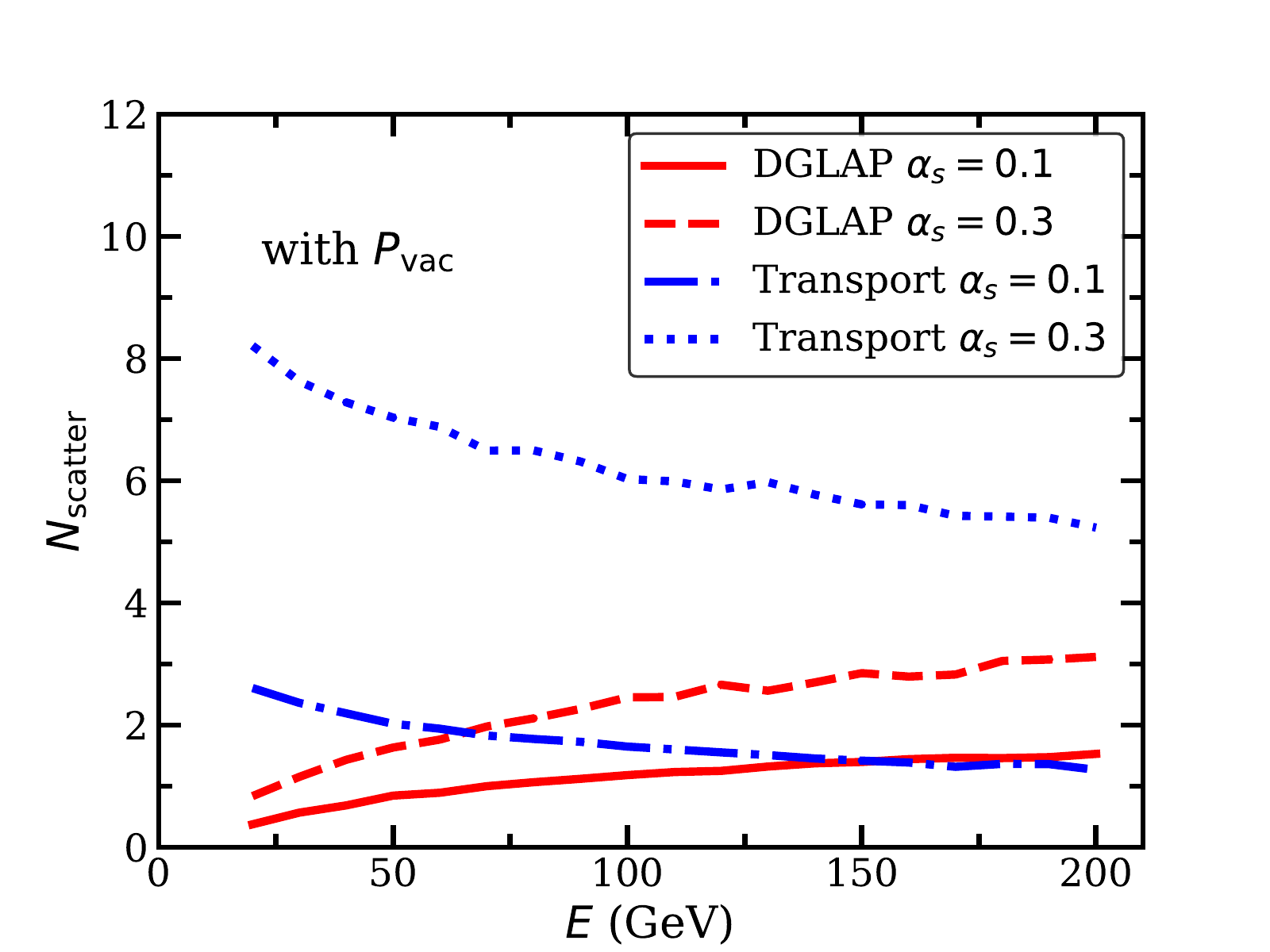}
        \caption{(Color online) The mean number of scatterings experienced by leading partons within the high virtuality (DGLAP) versus the low virtuality (transport) stages, with vacuum splitting function used in \textsc{Matter} simulation.}
        \label{fig:plot-average_Nscatter_g_vac}
\end{figure}

To better quantify the parton-medium interaction, we estimate the average number of scatterings for the leading partons within the two stages in Fig.~\ref{fig:plot-average_Nscatter_g_vac}. We track the path of each leading parton as it traverses the medium, and integrate its scattering rate $\Gamma$ along the path before and after its virtuality reaches $Q_0^2$. As shown in Fig.~\ref{fig:plot-average_Nscatter_g_vac}, this number of scatterings increases with the parton energy within the high virtuality (DGLAP) stage, while decreases within the low virtuality (transport) stage. 
Although increasing $\alpha_\mathrm{s}$ shortens the time inside the high virtuality (DGLAP) stage (as shown in Fig.~\ref{fig:plot-average_t_g_vac}), it increases the scattering rate and thus raises the scattering numbers in both DGLAP and transport stages. Note that on average, an energetic parton is able to scatter with the medium at least once while its virtuality is above $Q_0^2$. This further motivates the introduction of the medium-induced splitting additively within the DGLAP evolution equation for parton showers in relativistic heavy-ion collisions.


\section{NLO contributions and DGLAP evolution equation}
\label{sec:NLO}

In this section, we discuss how to include medium modification into the DGLAP evolution equation for parton showers at \emph{high virtualities}. 
We emphasize high virtuality as the final outcome will be different from that deduced in earlier references for a medium modified evolution equation~\cite{Guo:2000nz,Wang:2001ifa,Majumder:2009zu}, in particular, the medium modified portion will be found to be collinear divergent.

We start with revisiting the evolution equation in vacuum, where one considers the factorization of singular and finite terms from higher order calculations into the redefinition of the leading order fragmentation functions. Absorbing terms at all orders leads to the leading order resummed scale dependent fragmentation functions. We then consider the effects of splitting in a dense medium with one or less scattering per emission. This leads to the medium modified splitting kernel, which is also divergent in the limit of large virtuality. In the end, we extend the DGLAP evolution equation to incorporate contributions from both vacuum and medium-induced splitting functions, for studying medium effects on parton splittings at high virtualities.


\subsection{Vacuum DGLAP evolution equation}
\label{subsec:vacuum}

We begin by reminding the reader of the process of factorization of mass singularities into fragmentation functions in vacuum. While this can be carried out in several different schemes, we will carry this out in the momentum or virtuality cut-off scheme, which yields the most physical picture of the factorization procedure. 
Consider the case of an \epem \,event in vacuum with a photon invariant mass of $Q$. We define the coordinate system such that the virtual photon is static, with momentum $(Q,0,0,0)$, the quark defines the $-z$ direction, and the anti-quark the $+z$ direction, at leading order. In a purely partonic theory, one may express the cross section to produce a quark with light-cone momentum $q^- = Q/\sqrt{2} \equiv Q^-$, carrying a fraction $z = \sqrt{2} q^-/Q $ as, 
\bea
\frac{d \sigma}{ dz } = \sigma_0 D (z), 
\eea
where, $\sg_0$ is the leading order cross section for \epem \,annihilation to a $q\bar{q}$ pair, and $D(z) = \kd(1-z)$ is a quark to quark fragmentation function.

At Next-to-Leading Order (NLO), one obtains both a real ($\geq 0$) and a virtual ($\leq 0$) contribution. In the equation below, we include the leading logarithmic portion of the virtual correction along with the LO part, 
\bea
\frac{d \sigma}{dz} &=& \sigma_0 \left[ \delta(1 - z) 
\left\{1 -  \frac{\A_\mathrm{s}}{2\pi}  \int\limits_0^1 dy \int\limits_0^{Q^2y(1-y)} \frac{dl_\perp^2}{l_\perp^2}P(y) \right\} \right.  
																						\label{NLO_sigma} \\
&+&\left.  \frac{\A_\mathrm{s}}{2\pi}  \int\limits_z^1 \frac{dy}{y} \int\limits_0^{Q^2y(1-y)} \frac{d l_\perp^2}{l_\perp^2}  P(y) 
\kd \left( 1 - \frac{z}{y} \right) 
+  \frac{\A_\mathrm{s}}{2\pi} \int\limits_z^1 \frac{dy}{y} f(y) \kd \left( 1 - \frac{z}{y}\right) \right] \nn \\
&=& \sigma_0 \left[  \kd(1-z) + \frac{\alpha_\mathrm{s}}{2\pi} \int_0^{Q^2} \frac{d \mu^2}{\mu^2} \int_z^1 \frac{dy}{y} P_{+}(y) 
\kd \left( 1 - \frac{z}{y} \right)  + \frac{\A_\mathrm{s}}{2\pi} \int_z^1 \frac{dy}{y} f(y) \kd \left( 1 - \frac{z}{y} \right)  \right]. \nn
\eea
In the equation above, $P(y) = C_F \frac{1-y^2}{1-y}$. 
The virtual correction, the second term in curly brackets above, does not actually terminate at $l_\perp^2 = Q^2y(1-y)$, where $Q$ is the mass of the virtual photon. The portion from $Q^2 y(1-y)$ to $\infty$ is absorbed in a UV renormalization of the coupling, thereby giving $\A_\mathrm{s}$ a scale dependence, once higher order corrections are included. 
Also note that the upper limit of $l_\perp^2$ even in the real emission contribution is actually $Q^2y (1-y)$. To remove $y$ dependence in the limits of integration, we change variables to $\mu^2 = l_\perp^2 /[y(1-y)]$ in the last line.  This also allows us to reverse the order of integration.
In the last line of the above equation, 
we also combine the real and virtual terms to define the ($+$)-function [$P_{+} (y)$], where the infrared divergence 
is cancelled between these terms.  
The last term in the last line above, involving the function $f(y)$ represents the coefficient function, i.e., terms that do not possess the collinear singularity.

The NLO cross section has a collinear divergence which can be subtracted and absorbed into a redefinition of the scale dependent fragmentation function,
\bea
D(z,\mu_F^2) = D(z) + \frac{\alpha_\mathrm{s}}{2\pi} \int_0^{\mu_F^2} \frac{d \mu^2}{\mu^2} \int_z^1 \frac{dy}{y} P_{+} (y)
D \left( \frac{z}{y} \right) + \frac{\alpha_\mathrm{s}}{2\pi}  \int_z^1 dy f_D (y) D \left( \frac{z}{y} \right). 
																						\label{scale_dependent_D}
\eea
In this redefinition, 
a portion $f_D$ of the coefficient function $f$ is absorbed into the definition of the the scale dependent fragmentation function.
The choice of $f_D$ defines a renormalization scheme. In the equation above, the divergent integral $\int_0^{\mu_F^2} d\mu^2/\mu^2$
can be evaluated using either dimensional regularization, massive gluon scheme, or virtuality cut-offs. In each case, the function $f_D$ is different. 
We do not list out the various values of $f_D$ here, as we will show that it plays no role in the eventual evolution equations of the fragmentation functions, which 
are scheme independent. 
Including these terms changes the cross section to, 
\bea
\frac{d \sigma}{dz} &=& \sigma_0 D(z,\mu_F^2) 
+ \sigma_0 \frac{\A_\mathrm{s}}{2\pi} \left[ \int\limits_{\mu_F^2}^{Q^2} \frac{d\mu^2}{\mu^2}  
\int\limits_z^1 \frac{dy}{y} P_{+}(y) D \left( \frac{z}{y} \right) 
+  \int\limits_z^1 \frac{dy}{y} \left\{ f(y) - f_D(y)  \right\} D \left( \frac{z}{y} \right) \right] \nn \\
&=& \sigma_0 D(z,\mu_F^2) 
+ \sigma_0 \frac{\A_\mathrm{s}}{2\pi}  \int_z^1 \frac{dy}{y} \left[ \log\left(\frac{Q^2}{\mu_F^2} \right)  P_{+}(y) 
+ f(y) -f_D(y) \right] D \left( \frac{z}{y} \right) \nn \\ 
&=& \sigma_0 D(z,\mu_F^2) 
+ \sigma_0 \frac{\A_\mathrm{s}}{2\pi}  \left[ \log\left(\frac{Q^2}{\mu_F^2} \right)  P_{+} 
+ \Delta f \right] * D(z). 																\label{NLO_sigma_with_D_mu}
\eea
In the last line we have introduced the convolution notation of Feynman, 
\bea
P * D(z) = \int_z^1 \frac{dy}{y} P(y) D\left( \frac{z}{y} \right) = \int_z^1 \frac{dy}{y} P\left( \frac{z}{y} \right) D(y).
\eea
In Eq.~\eqref{NLO_sigma_with_D_mu}, we have also defined the remnant portion of the coefficient function, $\Delta f(y) =  f(y) - f_D(y)$, 
which is finite and not absorbed into the above  
redefinition of the scale dependent fragmentation function, $D(z,\mu_F^2)$. The reader will have noted that we are only considering the evolution of the 
non-singlet (NS) quark fragmentation function. This is merely for brevity, and to focus on the generalization to the medium modified evolution equation. 
All our results are straightforwardly generalized for the coupled singlet and gluon fragmentation functions. 

At NNLO, we obtain the higher order terms that have to be absorbed into the scale dependent fragmentation function, 
\bea
\delta D_{NNLO}(z, \mu_F^2) &=& \left(\frac{\alpha_\mathrm{s}}{2\pi} \right)^2 
\left[  \int_z^1 \frac{dy}{y}  \int_0^{\mu_F^2} \frac{d\mu^2}{\mu^2} P_+(y)  + f_{D} (y) \right] \nn \\
&\times& \left[ \int_{\frac{z}{y}}^1 \frac{dy_1}{y_1} \int_0^{\mu^2} \frac{d\mu_1^2}{\mu_1^2}P_+(y_1) + f_{D}(y_1) \right]
D\left(  \frac{z}{yy_1}\right)  \nn\\
&=& \frac{1}{2!} \left\{    \frac{\A_\mathrm{s}}{2\pi} \left( f_{D} + \int_0^{\mu_F^2} \frac{d\mu^2}{\mu^2} P_+ \right)  \right\}^2   *  D(z).  
\eea
The first two lines of the equation above are written in the limit of strong ordering of transverse momenta or virtualities, i.e. $\mu^2 \gg \mu_1^2$. 
The largest logarithmic corrections, order by order, are obtained in this limit. 
Resumming all orders, we obtain, 
\bea
D(z,\mu_F^2) &=& e^{  \frac{\A_\mathrm{s}}{2\pi} \left( f_{D} +  \int\limits_0^{\mu_F^2} \frac{d\mu^2}{\mu^2} P_+ \right)} * D(z),
\eea
which on differentiation yields the DGLAP evolution equation, 
\bea
\mu_F^2\frac{\prt D(z,\mu_F^2)}{\prt \mu_F^2} = \frac{\A_\mathrm{s}}{2\pi} \int_z^1 \frac{dy}{y} P_+(y) D\left(\frac{z}{y},\mu_F^2 \right) .     \label{vac_DGLAP}
\eea
As expected, the scheme dependent functions do not appear in the evolution equation above, and are only present in the definition of the fragmentation functions. 

In a real experiment, a hard parton is formed in a hard scattering at a light cone location $t_0^-$ with a light-cone momentum $q^-$; we have defined the direction of propagation of the 
parton as the $-z$ direction and assume it is formed at $x_\perp=0$, thus only the negative light cone location is relevant. 
The vacuum fragmentation functions do not depend on the location of formation or the light-cone energy of the parton that fragments. However, strictly adhering to the position and energy of the fragmenting parton, we could restate the evolution equation as, 
\bea
\mu_F^2\frac{ \left. \prt D(z,\mu_F^2, t_0^-) \right|_{q^-} }{\prt \mu_F^2} = \left.\frac{\A_\mathrm{s}}{2\pi} \int_z^1 \frac{dy}{y} P_+(y) D\left(\frac{z}{y},\mu_F^2 , t_0^- + \tau_F^-\right) \right|_{yq^-} .
														\label{vac_evol_q_tau}
\eea
The Eq.~\eqref{vac_evol_q_tau} above, states that the fragmentation function of a 
parton with light-cone momentum $q^-$ formed at the location $t_0^-$, changes as the 
virtuality of the parton $\mu_F^2$ is increased. This takes place as the parton splits within 
its formation time $\tau_F^- = 2 q^-/\mu_F^2$ by radiating away a fraction $1-y$ of its momentum fraction, and then 
fragmenting with light cone momentum $yq^-$ at a distance beyond $t_0^- + \tau_F^-$.


\subsection{Medium-modified DGLAP evolution equation}
\label{subsec:medium}

Consider the process highlighted in the preceding subsection, \epem\, annihilation leading to the formation of a back-to-back quark-antiquark jet, taking place within a dense medium. 
For simplicity, let us assume that the medium is static and $\hat{q}$ is a constant, as a function of position.
We now include terms from single re-scattering in the medium.  The cross section at NLO, the equivalent of Eq.~\eqref{NLO_sigma} but with a single re-scattering in a medium, can be expressed as: 
\bea
\frac{d \sigma}{dz} &=& \sigma_0 \left[  \kd(1-z) + \frac{\alpha_\mathrm{s}}{2\pi} \int\limits_0^{Q^2} \frac{d \mu^2}{\mu^2} \int\limits_z^1 \frac{dy}{y} P_{+}(y)
\kd \left( 1 - \frac{z}{y} \right)  + \frac{\A_\mathrm{s}}{2\pi} \int\limits_z^1 \frac{dy}{y} f(y) \kd \left( 1 - \frac{z}{y} \right)  \right. 
																				\label{NLO_sigma_rescatter}\\
&+& \left. \frac{\alpha_\mathrm{s}}{2\pi} \int\limits_0^{Q^2} \frac{d \mu^2}{\mu^2} \int\limits_z^1 \frac{dy}{y} 
\int\limits_{t_0^-}^{t_0^- + \tau^-} d \zeta^- \left\{P(y) \frac{\hat{q}(\zeta^-)}{\mu^2 y (1-y)} \right\}_+ \left\{ 2 - 2 \cos \left( \frac{ \mu^2 \zeta^-}{ 2q^-}\right) \right\}
 \kd \left( 1 - \frac{z}{y} \right) \right] \nn.
\eea
In the equation above, we have picked the ($-$)-light-cone direction as the direction of propagation of the hard quark in the dense medium. 
The large light-cone momentum of the quark is $q^-$, the other light cone momentum is 
$q^+ = \mu^2/(2 q^-)$. The process starts at the light-cone location $t_0^-$ ($t_0^-$ can be chosen to be 0 without loss of generality) and engenders a medium modification to the point $t_0^- + \tau^-$, where 
$\tau^- = 1/q^+$, or the formation time of the radiation. Integrating out the length $\zeta^-$ is possible once the functional dependence of $\hat{q}$ on $\zeta^-$ is known. 
Assuming $\hat{q}$ to be a constant and $t_0^- = 0$ for simplicity, we obtain, the medium modified term [last line of Eq.~\eqref{NLO_sigma_rescatter}] as, 
\bea
\frac{\alpha_\mathrm{s}}{2\pi} \int\limits_0^{Q^2} \frac{d \mu^2}{\mu^2} \int\limits_z^1 \frac{dy}{y} \left\{P(y) \frac{ 2\hat{q} }{\mu^2 y (1-y)} \right\}_+ \frac{2 q^-}{\mu^2} \left[1 - \sin(1)\right],
\eea
which diverges quadratically as $\mu^2 \ra 0$. If instead $\hat{q}$ varies as $1/\zeta^-$ or $1/{\zeta^-}^2$, this leads to a linear or a logarithmic infra-red (collinear) divergence with respect to $\mu^2$, respectively. Note that in this work, when $\hat{q}$ is written together with light cone variables, e.g. $\zeta^-$ and $\tau^-$, it is also defined in terms of light cone coordinates. 
Otherwise, $\hat{q}$ is defined in terms of Minkowski coordinates [see Eq.~\eqref{qhat-def}].

We point out that the medium-dependent piece, in the last line of Eq.~\eqref{NLO_sigma_rescatter} above, 
depends via integration limit on the hard scale $Q^2$. This is also the case with the vacuum emission term in the first line of Eq.~\eqref{NLO_sigma_rescatter}. 
In Eqs.~\eqref{NLO_sigma} and \eqref{scale_dependent_D}, the collinear divergence in the vacuum emission term was split off from the $d\mu^2$ integration and absorbed into a re-definition of the scale dependent fragmentation function. The same process will be followed for the combination of divergences from the vacuum and medium dependent pieces. 
We thus split both integrations at the factorization scale,
\bea
\int_0^{Q^2} \frac{d \mu^2}{\mu^2} = \int_0^{\mu_F^2} \frac{d\mu^2}{\mu^2} + \int_{\mu_F^2}^{Q^2} \frac{d\mu^2}{\mu^2}.
\eea
The integrals from $0$ to $\mu_F^2$ can be absorbed into a new object, the medium modified fragmentation function. 
While not explicitly expressed, this new object, depends on specifics of the medium, such as the actual density profile, 
the point where the actual parton was created $t_0^-$, the direction in which it propagates $\hat{n}$, as well as its light-cone energy $q^-$:

\bea
\left.D(z,\mu_F^2,t_{0}^{-})\right|_{q^-} &=& D(z) + \frac{\alpha_\mathrm{s}}{2\pi}  \int_0^{\mu_F^2} \frac{d \mu^2}{\mu^2} \int_z^1 \frac{dy}{y} \bigg\{ P_{+}(y) 
                                              \label{NLO_med_mod_frag_func} \\ 
&+& \left. 
\left[P(y) \frac{\hat{q}(\zeta^-)}{\mu^2 y (1-y)} \right]_+ \int_{t_0^-}^{t_0^- + \tau^-} \!\!\!\!d \zeta^-  \left[ 2 - 2 \cos \left( \frac{ \mu^2 \zeta^-}{ 2q^-}\right) \right] \right\}
 D \left( \frac{z}{y} \right). \nn
\eea
In the equation above, $\tau^{-} = 2 q^{-}/\mu^{2}$ and we have suppressed the writing of the finite coefficient $f_D$ which is absorbed in the medium modified fragmentation function. 

	\begin{figure}
	\centering
	\begin{minipage} [l] {0.45\textwidth}
	\includegraphics[width=\textwidth]{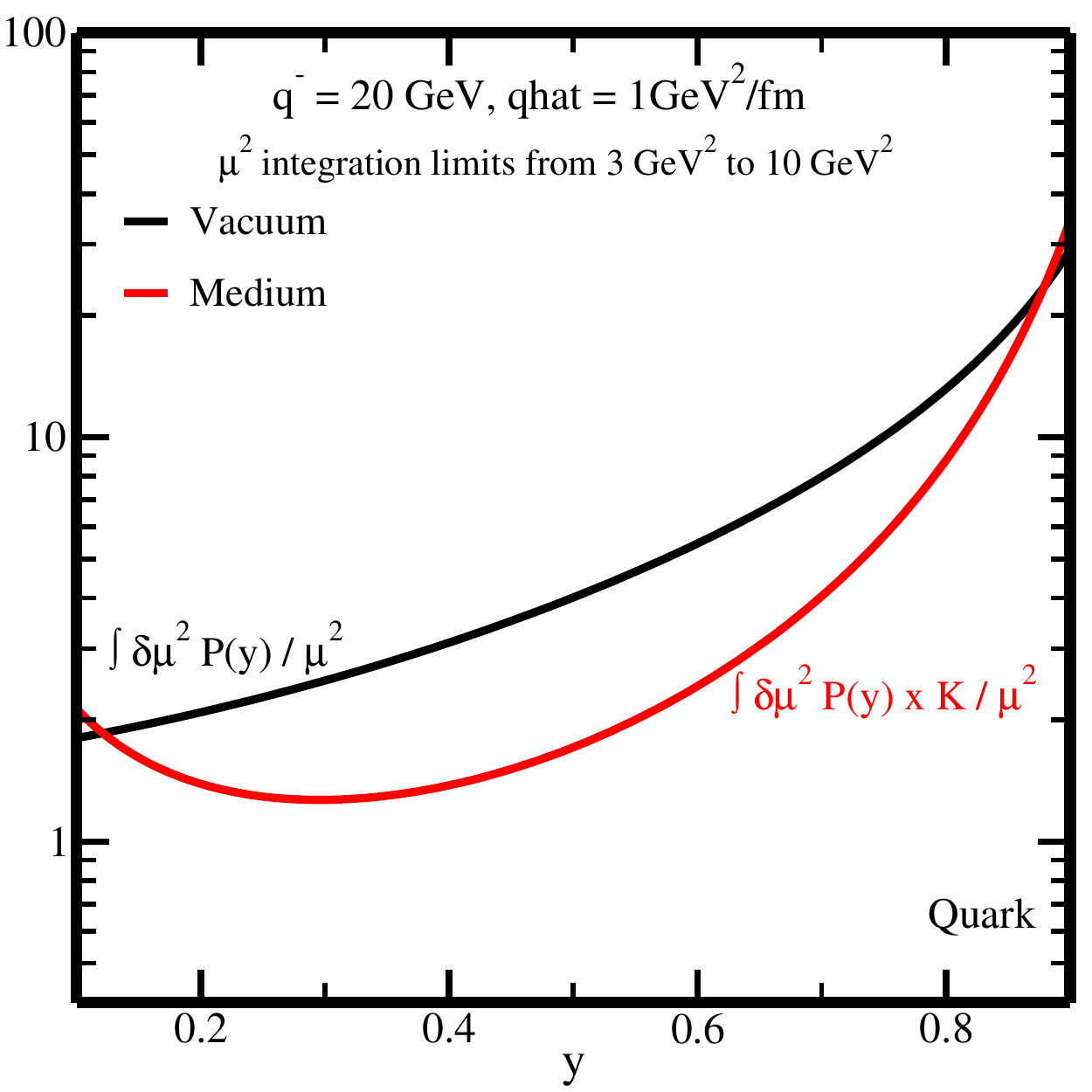}
	\end{minipage}
        \hspace{0.05\textwidth}
	\begin{minipage} [r] {0.45\textwidth}
	\includegraphics[width=\textwidth]{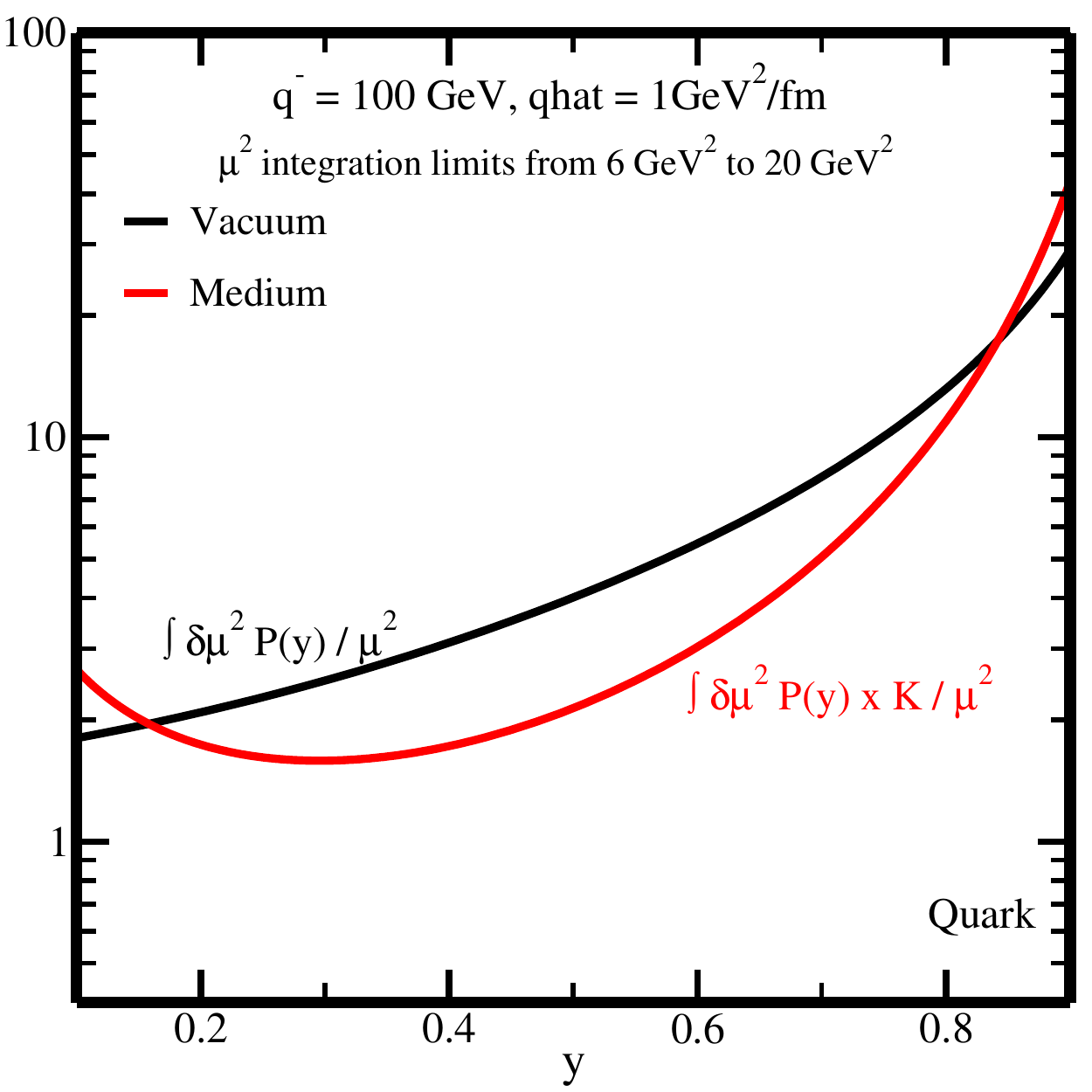}
	\end{minipage}
	\begin{minipage} [l] {0.45\textwidth}
	\includegraphics[width=\textwidth]{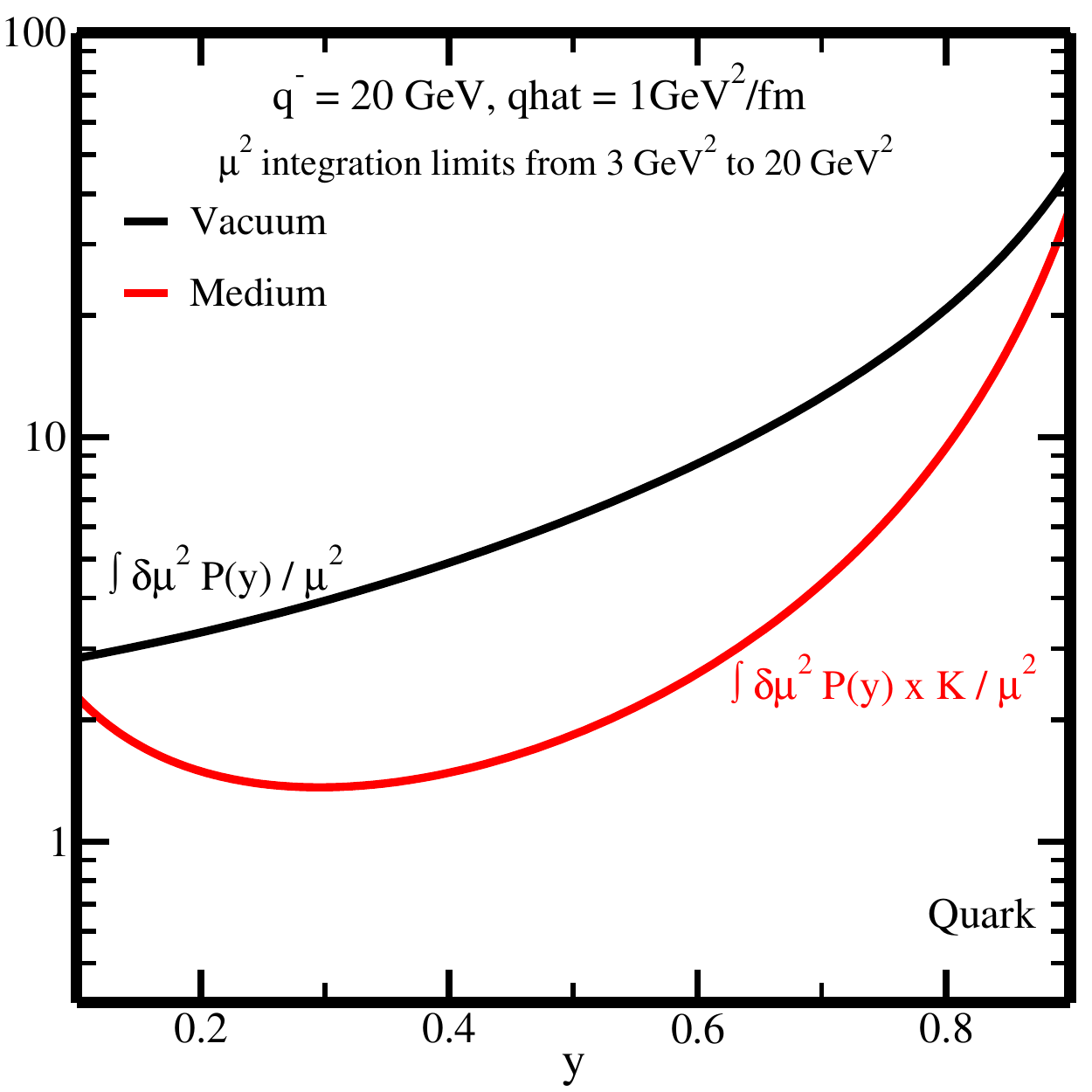}
	\end{minipage}
        \hspace{0.05\textwidth}
	\begin{minipage} [r] {0.45\textwidth}
	\includegraphics[width=\textwidth]{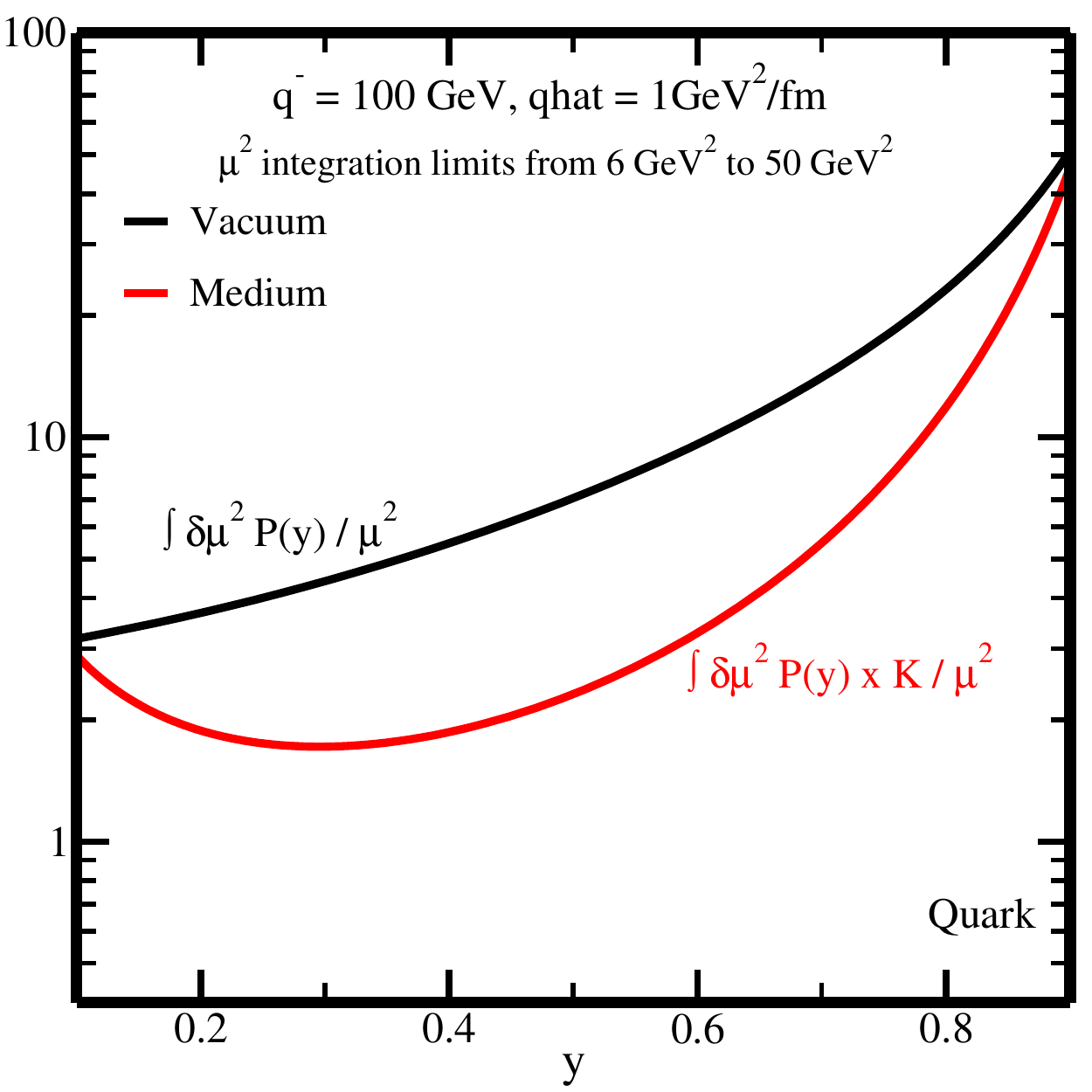}
	\end{minipage}
	\begin{minipage} [l] {0.45\textwidth}
	\includegraphics[width=\textwidth]{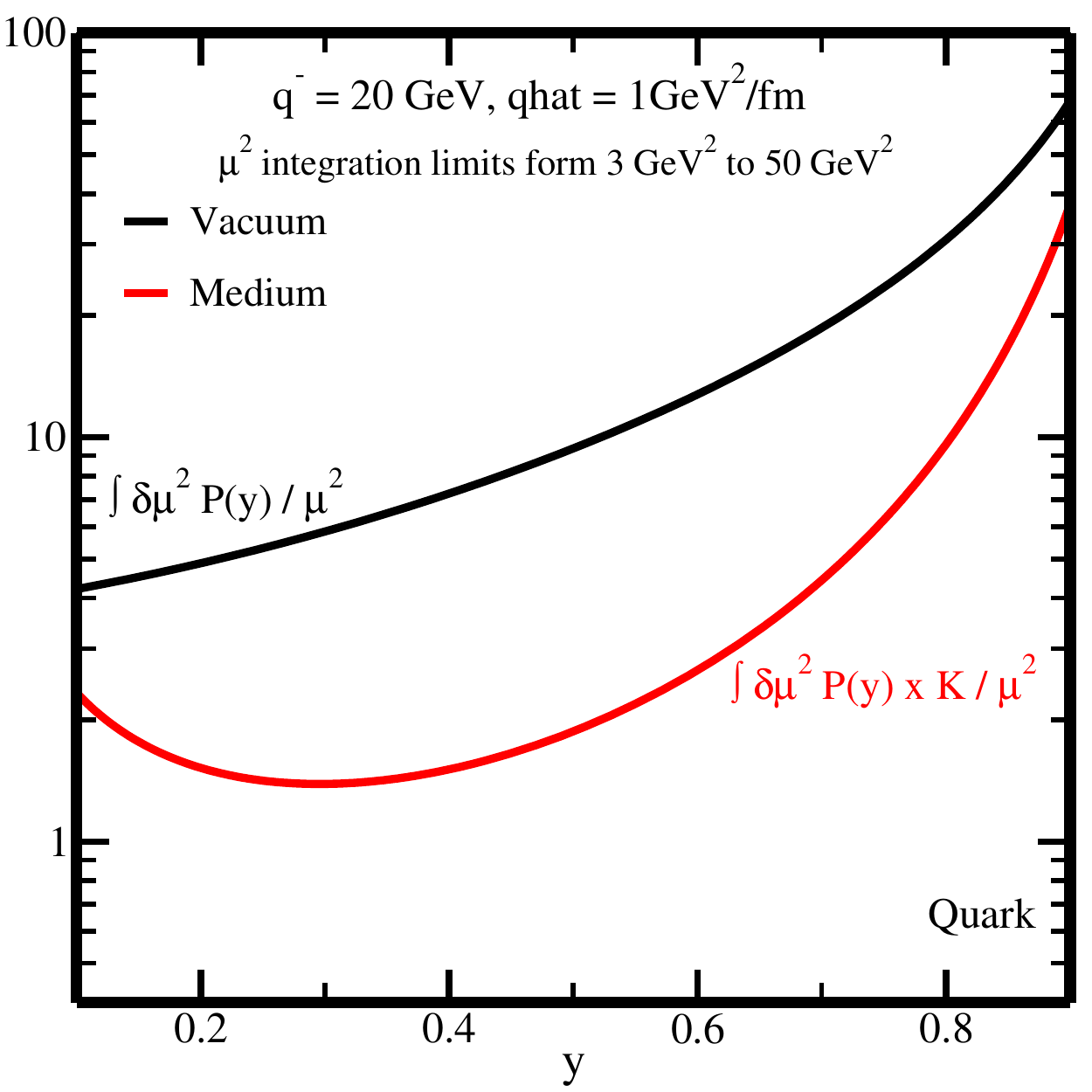}
	\end{minipage}
	\hspace{0.05\textwidth}
	\begin{minipage} [r] {0.45\textwidth}
	\includegraphics[width=\textwidth]{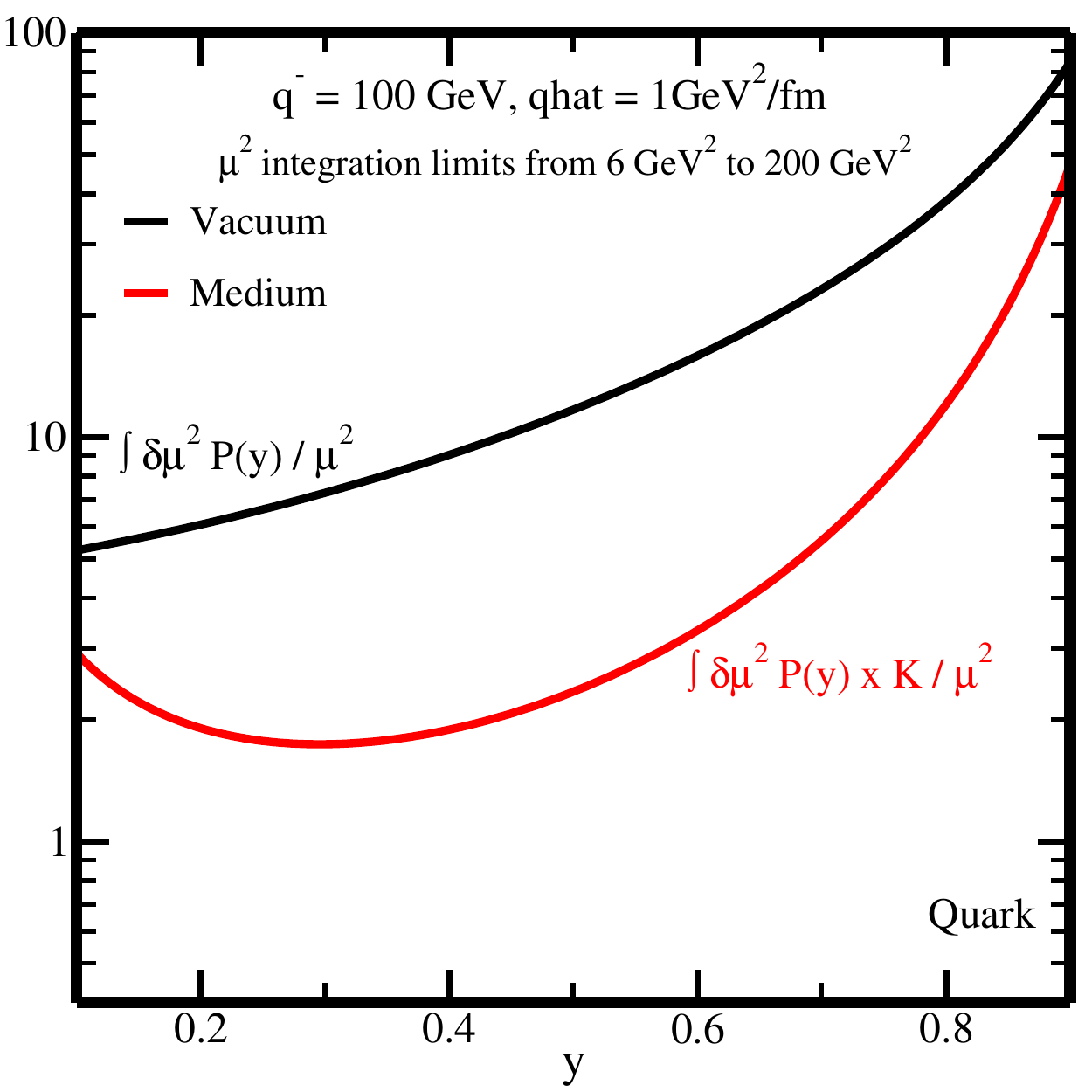}
	\end{minipage}
	\caption{Comparison between the medium modified kernel and the vacuum kernel of the fragmentation function from Eq.~\eqref{NLO_med_mod_frag_func}, for two different light-cone momentum $q^-=20$~GeV (left panels) and 100~GeV (right panels). The coefficient $\hat{q}$ is assumed to be constant and set to be 1~GeV$^2$/fm. The three plots for each energy 
	represent integrations over $\mu^2$ with the minimum set to $\mu^2_\mathrm{min} \simeq \sqrt{2 \hat{q} q^-}$. }
	\label{fig:MedK_Compare}
	\end{figure}

The new medium modified contribution is for the most part a small correction to the vacuum term. To illustrate this, in Fig.~\ref{fig:MedK_Compare}, we replace $D(z/y)$ with $\delta (1 - z/y)$ and plot the two contributions separately for quarks traversing a static medium with a fixed $\hat{q} = 1$~GeV$^2$/fm. We present plots for quark light-cone momenta $q^- = $~20~GeV (left panels) and 100~GeV (right panels). For each case the $\mu^2$ integral in Eq.~\eqref{NLO_med_mod_frag_func} is carried out over a range of values, with the minimum set to be $\mu^2 \simeq \sqrt{ 2 \hat{q} q^- }$, which represents the realistic minimum in virtuality down to which this formalism is applicable. As shown in Fig.~\ref{fig:MedK_Compare}, with an increasing virtuality scale (from the top to the bottom row), the relative contribution from the medium-modified part in Eq.~\eqref{NLO_med_mod_frag_func} decreases. For most of the available phase space, the medium-modified part can be viewed as a perturbation on top of the vacuum term.

Separating and absorbing the divergent terms in the scale dependent fragmentation functions, allows us to express the full cross section as, 
\bea
\frac{d \sigma}{d z} &=& \left. \sigma_0 D(z,\mu_F^2,t_{0}^{-})\right|_{q^-} 
+ \sigma_0 \frac{\A_\mathrm{s}}{2\pi}  \left[ \, \int\limits_{\mu_{F}^2}^{Q^{2}} \frac{d\mu^{2}}{\mu^{2}}  \left\{  P_{+} 
+  \!\!\!  \int\limits_{t_{0}^{-}}^{t_{0}^{-} + \tau^{-}} \!\!\!\!\! d\zeta^- \frac{\hat{q}}{\mu^2} K_+(\zeta^-)  \right\} \right] \!*\! D(z). 
\eea
Here, $K = P(y)[2 - 2 \cos( \mu^2 \zeta^-/ 2q^-)]/[y(1-y)]$, and we have suppressed the $\Delta f$ coefficient term that typically remains after the scheme dependent $f_D$ has been absorbed into 
the scale dependent fragmentation function.

One notes that the effective fragmentation function is now a function of the energy $q^-$ and the origin of the parton $t_0^-$. To highlight certain features of the ensuing evolution equation, we also write down the effective medium modified fragmentation function at NNLO, i.e., in the case of double emission with one scattering or less per emission:
%
%
	\begin{eqnarray}
	\left.D\left( z, \mu_F^2 , t_0^- \right)\right|_{q^-} &=& D\left( z \right) + \frac{ \alpha_\mathrm{s} } { 2 \pi } 
	\int_0^{\mu_F^2} \frac{ d \mu_0^2 } { \mu_0^2 } \int_z^1 \frac{ d y_0} { y_0 } P_{+}\left( y_0 \right) \\
	& \times & \left[ 1 + \int_{ t_0^- }^{ t_0^- + \tau_0^- } d \zeta_0^- \frac{ \hat{ q } \left( \zeta_0^- \right) } { \mu_0^2 y_0 \left( 1 - y_0 \right) }
	\left\{ 2 - 2 \cos \left( \frac{ \zeta_0^- } { \tau_0^- } \right) \right\} \right] 
 D\left( \frac{ z } { y_0 } \right) \nn \\ 
	&+&  \frac{ \alpha_\mathrm{s} } { 2 \pi } 
	\int_0^{\mu_F^2} \frac{ d \mu_0^2 } { \mu_0^2 } \int_z^1 \frac{ d y_0} { y_0 } P_{+}\left( y_0 \right) \nonumber\\
	& \times & \left[ 1 + \int_{ t_0^- }^{ t_0^- + \tau_0^- } d \zeta_0^- \frac{ \hat{ q } \left( \zeta_0^- \right) } { \mu_0^2 y_0 \left( 1 - y_0 \right) }
	\left\{ 2 - 2 \cos \left( \frac{ \zeta_0^- } { \tau_0^- } \right) \right\} \right] \nonumber\\ 
	& \times & \frac{ \alpha_\mathrm{s} } { 2 \pi } 
	\int_0^{\mu_0^2} \frac{ d \mu_1^2 } { \mu_1^2 } \int_{ z / y_0}^1 \frac{ d y_1} { y_1 } P_{+}\left( y_1 \right) \nonumber\\
	& \times & \left[ 1 + \int_{ t_0^- + \tau_0^-}^{ t_0^- + \tau_0^- + \tau_1^- } \!\!\!\!\!d \zeta_1^- \frac{ \hat{ q } \left( \zeta_1^- \right) } { \mu_1^2 y_1 \left( 1 - y_1 \right) }
	\left\{ 2 - 2 \cos \left( \frac{ \zeta_1^- } { \tau_1^{-} } \right) \right\} \right] 
	D \left( \frac{ z } { y_0 y_1 } \right) 
	+ ... \nn
	\end{eqnarray}
where $ \tau_0^- = 2 q^- / \mu_0^2 $ and $ \tau_1^- = 2 q^- y_0 / \mu_1^2 $. The ellipsis at the end is meant to highlight the existence of terms with 
an arbitrary number of emissions. 
One should note that given a strong ordering in virtualities $\mu_0^2 \gg \mu_1^2$, and no ordering in 
the momentum fraction $y_0 \sim y_1$, there is a strong ordering in formation times and angles of the radiated gluons. As a result, the order in scattering locations, $\zeta_1 \gg \zeta_0$ is justified. This indicates that in fact $\tau_1^- \gg \tau_0^-$ and so on. The formation times of the first set of radiations is very short, while those of the successive radiations tend to grow rapidly.  
As a result, the first set of medium modifications is small compared to the vacuum piece, but these tend to grow with each successive radiation. 
Below a certain $\mu^2$, the medium modification will indeed eclipse the vacuum term, and the high virtuality (single scattering per emission) formalism will become invalid. 
In an actual simulation, the partons that drop below the medium induced hard scale $Q_0^2 \sim \hat{q} \tau$ will transition from the high virtuality stage to a low virtuality transport 
stage where the vacuum piece is ignored and one deals with multiple medium scatterings. 
This can be enforced after resummation by constraining the lower limit of $\mu^2$ in the evolution from the hard scale $Q^2$ down to the medium induced scale $Q_0^2$.
In the above equation, we have also dropped the writing of the $f_D$  functions for brevity.

	Similar to the case in vacuum, we take the derivative of the fragmentation function with respect to the scale $ \mu_F^2 $. We also change the variables as $y_0 \rightarrow y$, $y_1 \rightarrow y_0$, $ \zeta_0^- \rightarrow \zeta^- $, $ \zeta_1^- \rightarrow \zeta_0^- $, $ \tau_1 \rightarrow \tau_0 $ and $\mu_1 \rightarrow \mu_0$ for convenience. Then we obtain,

\begin{eqnarray}
	\frac{ \partial \left. D\left( z, \mu_F^2 , t_0^-\right)\right|_{q^-} } { \partial \mu_F^2 } &=&  \frac{ \alpha_\mathrm{s} } { 2 \pi } 
	 \frac{ 1 } { \mu_F^2 } \int_z^1 \frac{ d y} { y } P_{+}\left( y \right) \\
	& \times & \left[ 1 + \int_{ t_0^- }^{ t_0^- + \tau_F^- } d \zeta^- \frac{ \hat{ q } \left( \zeta^- \right) } { \mu_F^2 y \left( 1 - y \right) }
	\left\{ 2 - 2 \cos \left( \frac{ \zeta^- } { \tau_F^- } \right) \right\} \right] 
	D\left( \frac{ z } { y } \right) \nn \\
	&+&  \frac{ \alpha_\mathrm{s} } { 2 \pi } 
	 \frac{ 1 } { \mu_F^2 } \int_z^1 \frac{ d y} { y } P_{+}\left( y \right) \nonumber\\
	& \times & \left[ 1 + \int_{ t_0^- }^{ t_0^- + \tau_F^- } d \zeta^- \frac{ \hat{ q } \left( \zeta^- \right) } { \mu_F^2 y \left( 1 - y \right) }
	\left\{ 2 - 2 \cos \left( \frac{ \zeta^- } { \tau_F^- } \right) \right\} \right] \nonumber\\ 
	& \times & \frac{ \alpha_\mathrm{s} } { 2 \pi } 
	\int_0^{\mu_F^2} \frac{ d \mu_0^2 } { \mu_0^2 } \int_{ z / y }^1 \frac{ d y_0} { y_0 } P_{+}\left( y_0 \right) \nonumber\\
	& \times & \left[ 1 + \int_{ t_0^- + \tau_F^- }^{ t_0^- + \tau_F^-  + \tau_0^- } \!\!\!\!d \zeta_0^- \frac{ \hat{ q } \left( \zeta_0^- \right) } { \mu_0^2 y_0 \left( 1 - y_0 \right) }
	\left\{ 2 - 2 \cos \left( \frac{ \zeta_0^- } { \tau_0^{-} } \right) \right\} \right] 
	D \left( \frac{ z / y } { y_0 } \right) + ... \nn
\end{eqnarray}

Resumming all higher order terms, with one or less scattering per emission, and differentiating with $\mu_F^2$ leads us to the medium modified DGLAP evolution equation as:

\begin{eqnarray}
\label{eq:med-DGLAP}
\frac{ \partial   D\left( z, \mu_F^2 , t_0^- \right)|_{q^-} } { \partial \mu_F^2 } &=&  \frac{ \alpha_\mathrm{s} } { 2 \pi } 
\frac{ 1 } { \mu_F^2 } \int\limits_z^1 \frac{ d y} { y }  \left[  P_+ ( y ) 
+  \left( \frac{ P(y) }{ y ( 1 - y ) }  \right)_{+} \right.  \label{med_evol_q_tau}  \\
  & \times & \left. \int\limits_{ t_0^- }^{ t_0^- + \tau_F^- } \!\!\!\! d \zeta^- \frac{ \hat{ q } \left( \zeta^- \right) } { \mu_F^2  }
\left\{ 2 - 2 \cos \left( \frac{ \zeta^- } { \tau_F^- } \right) \right\}  \right] 
 \left.  D \left( \frac{ z } { y },  \mu_F^2, t_0^- + \tau_F^- \right) \right|_{yq^-}   . \nn
\end{eqnarray}
One should take note of the locations of the effective fragmentation function on the right hand side of the equation above. As in the case of the vacuum in Eq.~\eqref{vac_evol_q_tau}, the evolution equation, mixes the fragmentation function at $t_0^-$ with functions at locations $t_0^- + \tau_F^- > t_0^-$. In the equation above $\tau_F^- = 2 q^- / \mu_F^2$. 
It is not entirely unexpected that the scale evolution of the medium modified fragmentation function will engender the mixing of fragmentation functions at two locations along the jet path. 
However, unlike the case in vacuum, the medium modified fragmentation function genuinely depends on location of formation and light cone momentum of the specific parton. 
This fact also makes the solution of the medium modified fragmentation function much more complicated.


\section{Nuclear modification within medium-modified DGLAP and transport}
\label{sec:med}

In the preceding section, the DGLAP evolution equation in vacuum was recast with the obvious energy and position dependence in Eq.~\eqref{vac_evol_q_tau}. 
These position and energy factors are typically ignored in the vacuum evolution equations as the vacuum fragmentation functions do not depend on position or energy. 
In the derivation of the medium modified evolution equations, these position and energy factors become essential as they control the amount of medium modification. 
The eventual medium modified fragmentation functions are thus position and energy dependent. 

In this section, we will generalize the DGLAP evolution equations to a Sudakov-like formalism to carry out a Monte-Carlo simulation of jet evolution. 
Since the main difference between Eq.~\eqref{vac_evol_q_tau} and Eq.~\eqref{med_evol_q_tau} is an additive term in the splitting function, this generalization is very straightforward, and can be carried out in exactly the same way as for the case in vacuum.

\subsection{Evolution time and number of scatterings}
\label{subsec:nScatt}

It has been justified in the previous section that medium modification to highly virtual partons can be taken into account by introducing the medium-induced splitting function into the DGLAP evolution equation. Compared to the vacuum scenario, the medium-modified fragmentation function in Eq.~(\ref{eq:med-DGLAP}) not only depends on the momentum fraction $z$ and parton scale $\mu_F^2$, but also depends on the path of each parton through the medium, which further depends on the production location and momentum of the parton. This complicates the numerical evaluation of Eq.~(\ref{eq:med-DGLAP}), especially within a dynamically evolving QGP medium, and thus a full Monte-Carlo simulation becomes a more natural choice.

To investigate the medium effects on parton splittings within the high virtuality (DGLAP) stage, we introduce the medium-induced contribution to the splitting function in our \textsc{Matter} event generator, or Eq.~(\ref{eq:sudakov}), as follows \begin{equation}
\label{eq:totP}
P(y,\tilde{t})=P^\mathrm{vac}(y)+P^\mathrm{med}(y,\tilde{t}),
\end{equation}
where the medium-induced part reads
\begin{align}
\label{eq:medP}
P^\mathrm{med}&(y,\tilde{t})=\frac{P^\mathrm{vac}(y)}{y(1-y)\tilde{t}}\int\limits_0^{\tau_F^-} d\zeta^- \hat{q}(r+\zeta^-)\Biggl[2-2\cos \left(\frac{\zeta^-}{\tau_F^-}\right) \Biggl]
\end{align}
as discussed above.
In Eq.~(\ref{eq:medP}), $\hat{q}$ is the gluon jet transport coefficient that is evaluated locally at $\vec{r}+\hat{n}\zeta^-$; $\tau^-_f=2p^-/\tilde{t}$ is the mean splitting time. While the jet partons are inside the dense nuclear matter, i.e. after the hydrodynamical medium commences ($\tau_0 = 0.6$~fm) and before the local temperature of the surrounding medium drops below $T_\mathrm{c} = 165$~MeV, both the vacuum and medium-induced parts contribute to the splitting function Eq.~(\ref{eq:totP}). On the other hand, before partons enter or after they exit the QGP, only the vacuum splitting function contributes to parton showers. 

The validity of Eq.~(\ref{eq:med-DGLAP}) depends on the constraint that the medium-induced piece of splitting function contributes a correction to the vacuum piece. To respect this constraint, we require 
\begin{equation} 
\frac{1}{y(1-y)}\int\limits_{0}^{\tau_F^{-}} d \zeta^- \frac{ \hat{ q } \left( \zeta^- \right) } { \mu_F^2  }
\left[ 2 - 2 \cos \left( \frac{ \zeta^- } { \tau_F^{-} } \right) \right]  < 1.
\end{equation}
By assuming $\hat{q}$ is independent of location and taking $y=1/2$, this yields $\mu_F^2 \gtrsim 1.3\,\hat{q}\tau_F^-$. Thus, in this section, we set our lower boundary of virtuality scale in \textsc{Matter} simulation as $Q_0^2 = 1.3\,\hat{q}\tau_F^-$. We have verified that shifting this stopping scale of \textsc{Matter} evolution between $\hat{q}\tau_F^-$ and $2\,\hat{q}\tau_F^-$ has little effect on the numerical results we present below.

\begin{figure}[tbp]
        \centering
                \includegraphics[width=0.55\linewidth]{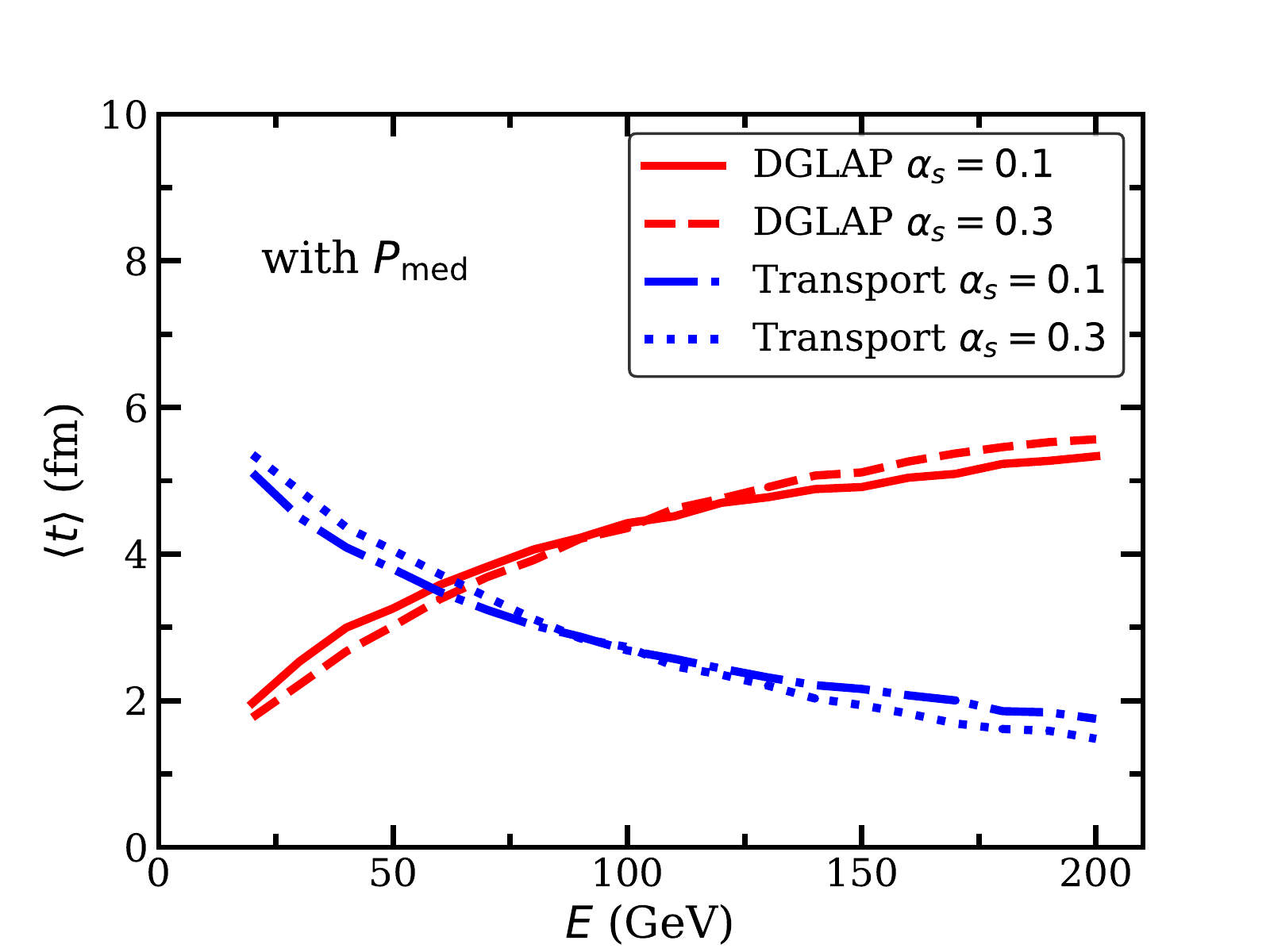}
        \caption{(Color online) The time spent by leading partons within high virtuality (DGLAP) versus low virtuality (transport) stages, with medium-modified splitting function used in \textsc{Matter}.}
        \label{fig:plot-average_t_g_med}
\end{figure}

\begin{figure}[tbp]
        \centering
                \includegraphics[width=0.55\linewidth]{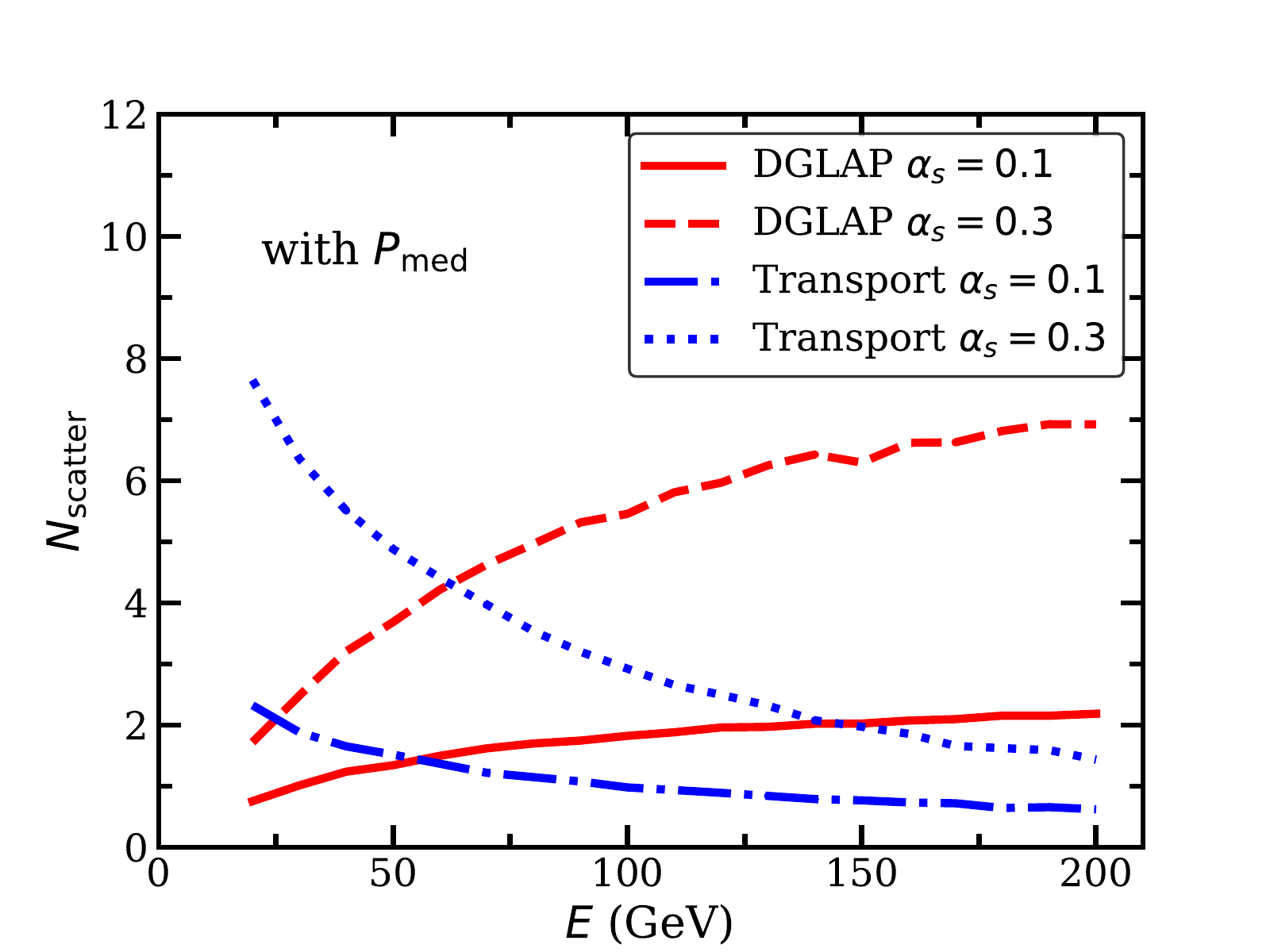}
        \caption{(Color online) The mean number of scatterings experienced by leading partons within the high virtuality (DGLAP) versus the low virtuality (transport) stages, with medium-modified splitting function used in \textsc{Matter}.}
        \label{fig:plot-average_Nscatter_g_med}
\end{figure}

Within this setup, we re-calculate the average propagation times and numbers of scatterings for leading gluons within the high virtuality (DGLAP) versus the low virtuality (transport) stages, as studied in Sect.~\ref{sec:vacscat}. Results with medium-modified splitting function in \textsc{Matter} simulation are presented in Figs.~\ref{fig:plot-average_t_g_med} and \ref{fig:plot-average_Nscatter_g_med}. By comparing Fig.~\ref{fig:plot-average_t_g_med} to Fig.~\ref{fig:plot-average_t_g_vac}, it is interesting to note that taking into account of parton-medium scattering significantly decelerates the decrease of parton virtuality -- the leading gluons spend much longer time during the high virtuality (DGLAP) stage when the medium-modified splitting function is applied, thus leaving a shorter time for the subsequent low virtuality (transport) evolution.

This is consistent with previous results presented in Ref.~\cite{Majumder:2014gda}, that partons lose virtuality more slowly inside a medium than in vacuum. In addition, although increasing $\alpha_\mathrm{s}$ raises the stopping scale $Q_0^2$ of the DGLAP evolution, it increases the parton-medium interaction as well and further slows down the virtuality decrease of partons. These two effects cancel and thus we find little $\alpha_\mathrm{s}$ dependence of the evolution time within the two stages here. Similarly, by comparing Fig.~\ref{fig:plot-average_Nscatter_g_med} to Fig.~\ref{fig:plot-average_Nscatter_g_vac}, one observes significantly more scatterings between the leading partons and the medium during the high virtuality (DGLAP) stage after the medium modification is introduced into the evolution equation. 
The above observations on time spent and scatterings encountered stipulates the need for medium modification to be included in the high virtuality phase in any real simulation of jets. 
This is carried out in the next subsection.


\subsection{Nuclear modification of single inclusive hadrons}
\label{subsec:RAA}

Throughout the preceding sections we have focussed on the relative contributions from medium modification in the high virtuality (DGLAP) stage 
versus that from the low virtuality (transport) stage of a hard jet. Our focus has been on the leading parton, the parton that contains a large fraction of the energy of the jet. In Sect.~\ref{sec:par} we argued that these high energy partons spend a considerable amount of time in the high virtuality 
(DGLAP) stage. 
In the preceding subsection, we demonstrated that the inclusion of medium modification factors leads to an extension of the time that the leading parton spends in the high virtuality stage. In this subsection, we explore the phenomenological implications of this. 
The fragmentation of the leading parton typically always contains the leading hadron.
 As a result, we investigate the contributions from DGLAP and transport evolutions to leading hadron observables in heavy-ion collisions.

 Parton showers are simulated using \textsc{Matter} when its virtuality is above $Q_0^2$. On the other hand, below $Q_0^2$, the parton evolution through the remaining path within the medium (with local temperature larger than $T_\mathrm{c}=165$~MeV) is simulated with the Linear Boltzmann Transport (\textsc{Lbt}) model~\cite{Cao:2016gvr,Cao:2017hhk,Luo:2018pto,He:2018xjv}. Details of combining \textsc{Matter} and \textsc{Lbt} into a multistage evolution model were discussed in Ref.~\cite{Cao:2017zih}.

Within \textsc{Lbt}, the parton-medium scattering is described by a set of rate equations. For the elastic process $ab\rightarrow cd$, the rate for a jet parton $a$ to scatter with a thermal parton $b$ from the medium is given by :
\begin{equation}
 \label{eq:rate2}
 \Gamma_a^\mathrm{el} = \sum_{b,c,d}\frac{\gamma_b}{2E_a}\int \prod_{i=b,c,d}d[p_i] f_b(\vec{p}_b) S_2(s,t,u) 
(2\pi)^4\delta^{(4)}(p_a+p_b-p_c-p_d)|\mathcal{M}_{ab\rightarrow cd}|^2,
\end{equation}
where  $d[p_i]=d^3p_i/[2E_i(2\pi)^3]$, and $\gamma_b$ and $f_b$ represent the spin-color degeneracy and thermal distribution of parton $b$ respectively. The leading-order matrix elements are utilized here with collinear  ($u,t\rightarrow 0$) divergence regulated by imposing the kinematic cut $S_2(s,t,u)=\theta(s\ge2\mu_\mathrm{D}^2)\theta(-s+\mu_\mathrm{D}^2\le t\le -\mu_\mathrm{D}^2)$, where $\mu_\mathrm{D}^2=6\pi\alpha_\mathrm{s}T^2$ is the Debye screening mass. The probability of elastic scattering of parton $a$ in each time step $\Delta t$ is thus $P_a^\mathrm{el}=\Gamma_a^\mathrm{el}\Delta t$. If an elastic scattering happens based on the Monte-Carlo method, the four-momentum of the out-going partons are then sampled based on the differential rate in Eq.~(\ref{eq:rate2}) without integration. Note that the same methods for simulating elastic scatterings have also been applied in \textsc{Matter} in this work, which slightly modifies the momentum of leading partons in virtuality-ordered splittings.

The inelastic scattering rate in \textsc{Lbt} at a given time $t$ is related to the average number of emitted gluons from the jet parton $a$ per unit time:
\begin{equation}
 \label{eq:gluonnumber}
 \Gamma_a^\mathrm{inel} (E_a,T,t) = \frac{1}{1+\delta_g^a}\int dydl_\perp^2 \frac{dN_g^a}{dy dl_\perp^2 dt},
\end{equation}
in which the $\delta_g^a$ term is imposed to avoid double counting for the $g\rightarrow gg$ process. The medium-induced gluon spectrum is also taken from the higher-twist energy loss formalism \cite{Guo:2000nz,Majumder:2009ge,Zhang:2003wk}:
\begin{eqnarray}
\label{eq:gluondistribution}
\frac{dN_g^a}{dy dl_\perp^2 dt}=\frac{2\alpha_\mathrm{s}(l_\perp^2) P^\mathrm{vac}_a(y)}{\pi l_\perp^4}\,\hat{q}\, {\sin}^2\left(\frac{t-t_i}{2\tau_F}\right),
\end{eqnarray}
which is consistent with the medium-induced splitting function used in Eqs.~(\ref{eq:med-DGLAP}) and (\ref{eq:medP}) considering that the transverse momentum of the emitted gluon with respect to its parent is related to the parent virtuality scale by $l_\perp^2 = y(1-y)t$. Here, $t_i$ denotes the production time of parton $a$. Similar to simulating the elastic process, we first use the integrated scattering rate Eq.~(\ref{eq:gluonnumber}) to determine whether medium-induced gluon is produced; and then use the differential spectrum Eq.~(\ref{eq:gluondistribution}) to sample its out-going momentum.

\begin{figure}[tbp]
        \centering
                \includegraphics[width=0.55\linewidth]{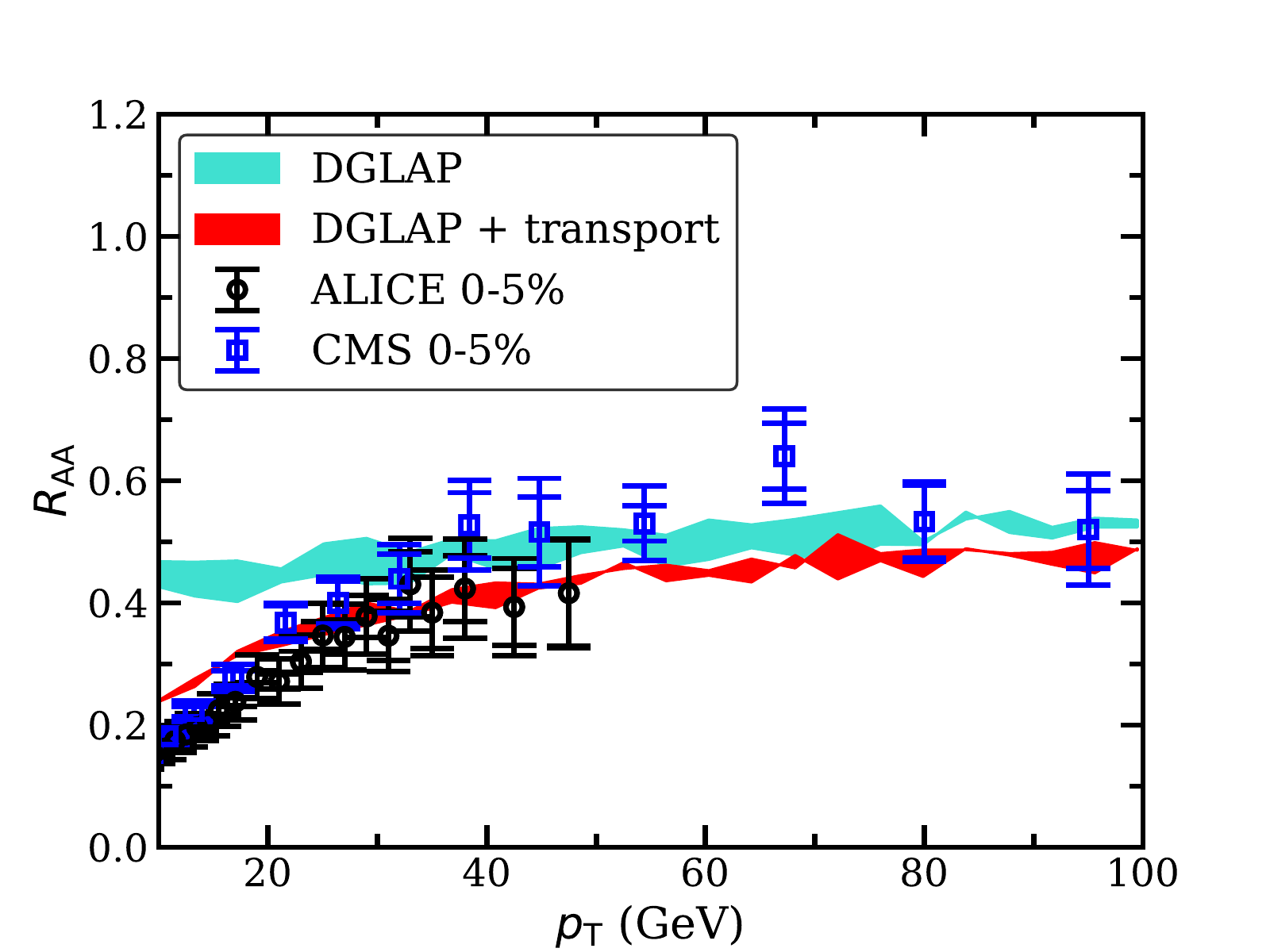}
        \caption{(Color online) Nuclear modification factor of single inclusive hadrons, contributed by DGLAP versus transport evolution.}
        \label{fig:plot-RAA-central}
\end{figure}

Combining \textsc{Matter} and \textsc{Lbt}, we calculate the nuclear modification factor of single inclusive hadrons in 2.76~ATeV central Pb-Pb collisions as presented in Fig.~\ref{fig:plot-RAA-central}. Here, energetic partons produced in initial hard collisions are initialized using the leading-order pQCD calculation in the momentum space, with the CTEQ parametrizations~\cite{Lai:1999wy} for parton distribution functions (PDF) and the EPS09 parametrizations~\cite{Eskola:2009uj} for the nuclear PDF modifications. These hard partons then evolve through our \textsc{Matter}+\textsc{Lbt} framework in which we fully track all parent and daughter partons. In vacuum (for proton-proton collisions), only \textsc{Matter} is applied, which evolves all partons to the minimum virtuality scale $t_\textrm{min} = 1~$GeV$^2$. With the presence of the medium, \textsc{Matter}+\textsc{Lbt} is applied. If partons exit the medium with virtuality larger than $t_\textrm{min}$, \textsc{Matter} vacuum shower is applied again until all partons approach $t_\textrm{min}$. At $t_\textrm{min}$, all partons are converted into hadrons using \textsc{Pythia 6} fragmentation~\cite{Sjostrand:2006za}. In this work, we adopt a minimum assumption of the $\hat{q}$ parametrization as $\hat{q}/T^3 = 2.0$ for quark (4.5 for gluon). This is equivalent to use Eq.~(\ref{eq:qhat}) to calculate $\hat{q}$ with $\alpha_\mathrm{s}=0.15$ for a $E=50$~GeV parton inside a $T=400$~MeV medium, and is around the lower boundary of the $\hat{q}$ values constrained by the JET Collaboration~\cite{Burke:2013yra}. Precise description of data and extraction of $\hat{q}$ is beyond the purpose of this work and will be reported separately.

In Fig.~\ref{fig:plot-RAA-central}, the DGLAP result is obtained by applying only \textsc{Matter} for medium modification of partons with virtuality above $Q_0^2$, while the DGLAP+Transport result is obtained by applying our full \textsc{Matter}+\textsc{Lbt} calculations. One observes that the medium-modified splittings in the high virtuality (DGLAP) stage alone already lead to significant amount of suppression of the hadron spectrum. At high $p_\mathrm{T}$, the hadron $R_\mathrm{AA}$ is only slightly further suppressed after invoking the transport evolution since high energy partons scatter with the medium mainly during the high virtuality (DGLAP) stage (as shown in Figs.~\ref{fig:plot-average_t_g_med} and \ref{fig:plot-average_Nscatter_g_med}). Lower momentum partons approach $Q_0^2$ faster and therefore engender a stronger nuclear modification during the transport stage. As a result, significant further suppression from the transport evolution is seen for the hadron $R_\mathrm{AA}$ at low $p_\mathrm{T}$. The combined nuclear modification of parton showers at high virtuality (DGLAP) and low virtuality (transport) stages, with a density and energy dependent boundary, naturally provides a reasonable description of the experimental data. 

The goal of the current effort is \emph{not} a comprehensive exploration and comparison with experimental data. 
The main goal is to argue, purely on theoretical grounds that if the core of high energy jets is modified by the medium, then this 
modification must have taken place while the leading parton is at a high virtuality $\mu^2 \gnsim \hat{q} \tau$. The modification of the 
leading parton is best reflected in the suppression of leading hadrons at high transverse momentum, and we have simply demonstrated 
that our formalism yields a consistent description of experimental data for leading hadron suppression.

\section{Summary}
\label{sec:summary}

In this paper, we have outlined the mechanism of medium modification of the hard core of a high transverse momentum (high-$p_\mathrm{T}$) jet. 
We argue this on both phenomenological grounds, where we observe a suppression in the leading hadron yield in a heavy-ion collision up to 
the highest $p_\mathrm{T}$. 
We also argued this on purely theoretical grounds, starting from a completely parametric considerations in Sect.~\ref{sec:par}. 
Two stages of energy loss of the leading parton were delineated: the high virtuality (DGLAP) stage where the virtuality of the 
parton $\mu^2 \gnsim \hat{q} \tau$, where $\hat{q}$ is the jet transport coefficient and $\tau$ is the lifetime of the parton 
(the more accurate relation for a space-time dependent $\hat{q}$ is $\mu^2 \gnsim \int_0^\tau d \zeta \hat{q} (r_i + \hat{n}\zeta) $). 
In Sect.~\ref{sec:vacscat} we demonstrated in a semi-realistic scenario with no energy loss in the high-virtuality stage that the leading 
parton spends a considerable amount of time in the high virtuality stage; the time spent increases with the energy of the parton. 
Estimating the number of scatterings engendered by the parton in this time, it becomes impossible to argue for no medium interaction in the 
high virtuality stage.  

In Sect.~\ref{sec:NLO}, we re-derived the medium modification of the DGLAP evolution equation for a highly virtual parton in a dense medium. 
We obtained that, in this limit, the medium modification is actually divergent and needs to be re-summed into a medium modified fragmentation function. The evolution of this fragmentation function gives the medium modified splitting function, which is then used in Sect.~\ref{sec:med} to first retest our semi-realistic estimate of time spent by the leading parton in the high virtuality stage. We found that the inclusion of medium modifications to the splitting function actually increased the time and number of scatterings engendered by a parton in the high virtuality stage. We then checked the effect of this on leading hadron suppression, the observable that should be most sensitive to 
the leading parton. Our results show a very good agreement with experimental data, further cementing the notion that there is indeed 
medium modification to parton evolution even when the virtuality of the parton is much larger than the medium scale $\hat{q} \tau$.

In this effort, we did not explore any evolution of the transport coefficient $\hat{q}$ itself with the scale of the parton~\cite{Kumar:2019uvu}. 
This will require extensive comparison with experimental data for jets and leading hadrons. This in turn will require 
the introduction of a scattering and recoil kernel which will have to be set up to yield the requisite $\hat{q}$. The number of 
parameters and data involved will necessitate a Bayesian analysis of the data. 
Such phenomenological extractions will have to be carried out as part of the JETSCAPE framework~\cite{Cao:2017zih,Putschke:2019yrg}.

\acknowledgments
We are grateful to the computing resource provided by the Open Science Grid (OSG). This work was supported in part by the U.S. Department of Energy (DOE) under grant number DE-SC0013460, and in part by the National Science Foundation (NSF) within the framework of the JETSCAPE collaboration under grant number ACI-1550300.

\bibliographystyle{JHEP}
\bibliography{SCrefs}

\end{document}